\documentclass[12pt,a4paper]{article}
\usepackage[T1]{fontenc}
\usepackage[latin1]{inputenc}
\usepackage{amsmath}
\usepackage{amssymb}

\usepackage{amsfonts,amsthm,euscript}
\usepackage{graphicx}

\usepackage[ref,nosort]{overcite} 

\makeatletter

\renewcommand\@biblabel[1]{$^{#1}$} 

\renewcommand{\thesection}{\Roman{section}}
\renewcommand{\@seccntformat}[1]{\csname the#1\endcsname.\hspace{0.5em}}

\renewenvironment{abstract}{%
  \begin{center}%
    {\bfseries \abstractname\vspace{-.5em}\vspace{\z@}}%
  \end{center}%
  \quotation}

\makeatother

\newtheorem{thm}{Theorem}
\newtheorem{lem}[thm]{Lemma}
\newtheorem{prop}[thm]{Proposition}
\newtheorem{cor}[thm]{Corollary}
\theoremstyle{remark}
\newtheorem*{rem*}{Remark}

\newcommand{\GG}{\ensuremath{\EuScript G}}
\newcommand{\KK}{\ensuremath{\EuScript K}}
\newcommand{\NN}{\ensuremath{\EuScript N}}
\newcommand{\LL}{\ensuremath{\EuScript L}}
\newcommand{\sS}{\ensuremath{\EuScript S}}
\newcommand{\TT}{\ensuremath{\EuScript T}}
\newcommand{\DD}{\ensuremath{\EuScript D}}
\newcommand{\HH}{\ensuremath{\EuScript H}}

\newcommand{\gq}{\ensuremath{\mathfrak q}}
\newcommand{\gf}{\ensuremath{\mathfrak f}}
\renewcommand{\gg}{\ensuremath{\mathfrak g}}
\newcommand{\gK}{\ensuremath{\mathfrak K}}
\newcommand{\gD}{\ensuremath{\mathfrak D}}

\newcommand{\dd}{\ensuremath{\,\mathrm{d}}}
\newcommand{\ii}{\ensuremath{\mathrm{i}}}
\newcommand{\ee}{\ensuremath{\mathrm{e}}}

\newcommand{\Dom}{\operatorname{Dom}}
\newcommand{\Ker}{\operatorname{Ker}}
\newcommand{\Mat}{\operatorname{Mat}}
\newcommand{\rank}{\operatorname{rank}}
\newcommand{\GL}{\operatorname{GL}}

\renewcommand{\Re}{\operatorname{Re}}
\renewcommand{\Im}{\operatorname{Im}}

\begin{document}

\title{On the Pauli operator for the Aharonov-Bohm effect with two
  solenoids}

\author{V.~A. Geyler\\
\emph{\normalsize Department of Mathematics, Mordovian State University}\\
\emph{\normalsize Bolshevistskaya 68, Saransk 430000, Russia}\\
\\
P. \v{S}\v{t}ov\'{\i}\v{c}ek\\
\emph{\normalsize Department of Mathematics, Faculty of Nuclear Science
}\\
\emph{\normalsize Czech Technical University}\\
\emph{\normalsize Trojanova 13, 120 00 Prague, Czech Republic}}

\date{{}}

\maketitle
\begin{abstract}
  \noindent We consider a spin-1/2 charged particle in the plane under
  the influence of two idealized Aharonov-Bohm fluxes. We show that
  the Pauli operator as a differential operator is defined by
  appropriate boundary conditions at the two vortices. Further we
  explicitly construct a basis in the deficiency subspaces of the
  symmetric operator obtained by restricting the domain to functions
  with supports separated from the vortices. This construction makes
  it possible to apply the Krein's formula to the Pauli operator.
\end{abstract}

\newpage

\section{Introduction}

The goal of the present paper is to provide a more detailed analysis
of the Pauli operator describing a spin-1/2 charged particle under the
influence of two Aharonov-Bohm (AB) fluxes \cite{AB}. We consider the
idealized setup when the magnetic fluxes are concentrated along two
parallel lines so that the problem effectively reduces to a
two-dimensional quantum system living in a perpendicular plane. In
what follows we call the intersection points of the fluxes with the
plane vortices.

To define the Pauli Hamiltonian with singular fluxes we use the
Aharonov-Casher decomposition \cite{AharonovCasher}. It makes it
possible to introduce the two diagonal components of the Pauli
operator corresponding to spin up and down as the unique selfadjoint
operators associated to appropriate quadratic forms. Since the
magnetic field vanishes outside of the vortices the two components of
the Pauli Hamiltonian as well as the spinless AB Hamiltonian are
selfadjoint extensions of the same symmetric operator. In the case of
one AB flux all the selfadjoint extensions are known to be defined by
appropriate boundary conditions at the vortex
\cite{PSLDabrow,AdamiTeta}. Thus our first goal was to distinguish the
boundary conditions defining the two components of the Pauli
Hamiltonian.

The second goal was to construct a basis in the deficiency subspaces
in the two-vortex case. In this case as well the two diagonal
components of the Pauli Hamiltonian and the spinless Hamiltonian are
selfadjoint extensions of a common symmetric operator. We show that
the deficiency indices of this symmetric operator are $(4,4)$. The
construction is based on the observation that the coefficients
$\psi(x)$ standing at singular terms in the asymptotic expansion in
the variable $x_{0}$ at a vortex of the Green function
$\GG_{z}(x,x_{0})$ belong to the deficiency subspace with spectral
parameter $z$.  Here we make use of the explicit knowledge of the
spinless two-vortex Green function $\GG_{z}(x,x_{0})$ \cite{PS:PLA}.

The next and final goal which naturally follows is a construction of
the two-vortex Green function for the Pauli Hamiltonian with the aid
of the Krein's formula. Even this problem is solved explicitly.
Surprisingly many features can be again derived from the asymptotic
analysis near a vortex.

The paper is organized as follows. In Section~\ref{sec:Preliminaries}
we summarize some basic facts and formulae concerning the spinless AB
Hamiltonian with one and two vortices. In
Section~\ref{sec:The-Pauli-Hamiltonian} we introduce the Pauli
operator with one and two AB fluxes and derive the boundary conditions
at a vortex defining the spin up and down components of the Pauli
Hamiltonian. In Section~\ref{sec:Asymptotic-behaviour} we provide a
basic asymptotic analysis near a vortex of functions from the
deficiency subspaces as well as that of the spinless Green function.
In Section~\ref{sec:Deficiency-subspaces} we construct a basis in the
deficiency subspaces. Section~\ref{sec:The-Kreins-formula} is devoted
to the application of the Krein's formula to our problem.

\section{Preliminaries. The AB Hamiltonian for a spinless particle
  \label{sec:Preliminaries}}

The AB Hamiltonian with one vortex and describing a spinless particle,
$H_{0}$, was introduced in Ref.~\citen{AB} and studied in a long
series of papers by many authors. For example, one can consult
Ref.~\citen{Ruijsenaars} for some mathematical details. It acts in
$L^{2}(\mathbb{R}^{2},\dd ^{2}x)$ and is nothing but the selfadjoint
operator associated to the closure of the positive quadratic form
\begin{equation}
  \int _{\mathbb{R}^{2}}\Bigg (\left|\left(\partial_{x_{1}}
      - \ii \, \frac{\alpha \, x_{2}}{\left|x\right|^{2}}\right)\!
    \varphi \right|^{2}+\left|\left(\partial _{x_{2}}
      +\ii \, \frac{\alpha \, x_{1}}{\left|x\right|^{2}}\right)\!
    \varphi \right|^{2}\Bigg )\dd ^{2}x\, ,
  \label{eq:QFormStdABHamilton}
\end{equation}
defined on the space of test functions
$\DD(\mathbb{R}^2\setminus\{0\})$. In other words, $H_0$ is the
Friedrichs extension of the corresponding symmetric operator with the
domain $\DD(\mathbb{R}^2\setminus\{0\})$. Owing to the gauge
equivalence we can assume that $\alpha \in (0,1)$.

We shall use the polar coordinates $(r,\theta )$ with the angle
$\theta\in(-\pi,\pi)$. This implies a cut along the negative $x_{1}$
half-axis. Sometimes it is convenient to apply the unitary operator
\[
(U_{\alpha }\varphi)(r,\theta) = \ee^{\ii\,\alpha\,\theta }
\varphi(r,\theta )
\]
and work with the unitarily equivalent operator
\[
H=U_{\alpha }H_{0}U_{\alpha }^{-1}.
\]
In particular this unitary transformation is useful when constructing
the Green function. This means that
\[
\Dom (H)=U_{\alpha }(\Dom (H_{0}))\, .
\]
Formally, as a differential operator,
\[
H=-\Delta \, .
\]
The domain of $H$ is determined by the boundary conditions at the
cut, namely
\begin{equation}
  \psi (r,\pi )=\ee ^{2\pi \, \ii \, \alpha }\psi (r,-\pi ),
  \textrm{ }\partial _{r}\psi (r,\pi )=\ee ^{2\pi \, \ii \, \alpha }
  \partial_{r}\psi (r,-\pi )\, .
  \label{eq:BCOneVortex}
\end{equation}
In addition, one should take care about boundary conditions at the
vortex. As analyzed in Refs.~\citen{PSLDabrow,AdamiTeta}, the domain
of $H_0$ is characterized by the boundary condition
$\varphi(0)=0$. Since
$\psi(r,\theta)=\exp(\ii\alpha\theta)\varphi(r,\theta)$ the same is
true for $\Dom(H)$, namely the boundary condition at the vortex reads
$\psi(0)=0$.

The generalized eigenfunctions of $H$,
\[
\left\{ \frac{1}{\sqrt{2\pi }}\,
  J_{\left|n+\alpha \right|}(k\, r)\, \ee ^{\ii (n+\alpha )\theta
  }\right\} _{k>0,\, n\in \mathbb{Z}},
\]
form a complete normalized set,
\[
\int _{0}^{\infty }J_{\nu }(k\, x)\, J_{\nu }(k\, y)\, k\dd
k=\frac{1}{x}\, \delta (x-y)\,.
\]
This makes it possible to write down the Green function and the
propagator as integrals,
\begin{equation}
\GG _{z}(r,\theta
  ;r_{0},\theta _{0})=\frac{1}{2\pi }\sum _{n\in \mathbb{Z}}\ee ^{\ii
    (n+\alpha )(\theta -\theta _{0})}\int _{0}^{\infty
  }\frac{J_{\left|n+\alpha \right|}(k\, r)\, J_{\left|n+\alpha
      \right|}(k\, r_{0})}{k^{2}-z}\, k\dd
  k\label{eq:GreenFceJJ}\end{equation} and\begin{equation} \KK
  _{t}(r,\theta ;r_{0},\theta _{0})=\frac{1}{2\pi }\sum _{n\in
    \mathbb{Z}}\ee ^{\ii (n+\alpha )(\theta -\theta _{0})}\int
  _{0}^{\infty }\ee ^{-\ii \, k^{2}t}J_{\left|n+\alpha \right|}(k\,
  r)\, J_{\left|n+\alpha \right|}(k\, r_{0})\, k\dd k\,.
\label{eq:PropagatorJJ}
\end{equation}
They are related by the Laplace transform,
\[ \GG _{z}(r,\theta
;r_{0},\theta _{0})=\int _{0}^{\infty }\ee ^{z\, t}\, \KK _{-\ii \,
  t}(r,\theta ;r_{0},\theta _{0})\dd t\, .
\]

Starting from (\ref{eq:PropagatorJJ}) one can derive the following
formula for the propagator\cite{PS:PLA},
\begin{eqnarray}
  \KK_{t}(r,\theta ;r_{0},\theta _{0}) & = & \left\{ \begin{array}{c}
      1\\
      \ee ^{2\pi \, \ii \, \alpha }\\
      \ee ^{-2\pi \, \ii \, \alpha }\end{array}
  \right\} \frac{1}{4\pi \, \ii \, t}\exp \!
  \Big (-\frac{1}{4\ii \, t}\left|x-x_{0}\right|^{2}\Big )
  \label{eq:FinalPropag}\\
  & & -\frac{\sin (\pi \, \alpha )}{\pi }\int _{-\infty }^{\infty
  }\frac{1}{4\pi \, \ii \, t}\exp \! \Big (-\frac{1}{4\ii \, t}\,
  R(s)^{2}\Big )\frac{\ee ^{-\alpha \, s+\ii \, \alpha \, (\theta
      -\theta _{0})}}{1+\ee ^{-s+\ii \, (\theta -\theta _{0})}}\,
  \dd s\,,\nonumber
\end{eqnarray}
where
\[
\left|x-x_{0}\right|^{2}=r^{2}+r_{0}^{2}-2r\, r_{0}\cos (\theta
-\theta _{0}),\textrm{ }R(s)^{2}=r^{2}+r_{0}^{2}+2r\, r_{0}\cosh (s),
\]
and the phase factor in front of the first term depends on whether
\[
\theta -\theta _{0}\in (-\pi ,\pi ),\textrm{ }(\pi ,2\pi )
\textrm{ or }(-2\pi ,-\pi ).
\]
The Laplace transformation results in a formula for the Green
function,
\begin{eqnarray}
  \GG _{z}(r,\theta ;r_{0},\theta _{0})
  & = & \left\{ \begin{array}{c} 1\\
      \ee ^{2\pi \, \ii \, \alpha }\\
      \ee ^{-2\pi \, \ii \, \alpha }
    \end{array}
  \right\} \frac{1}{2\pi }\, K_{0}\!
  \left(\sqrt{-z}\left|x-x_{0}\right|\right)\label{eq:FinalGreenFce}\\
  &  & -\frac{\sin (\pi \, \alpha )}{\pi }
  \int _{-\infty }^{\infty }\frac{1}{2\pi }\, K_{0}
  \Big (\sqrt{-z}\, R(s)\Big )\frac{\ee ^{-\alpha \, s
      +\ii \, \alpha \, (\theta -\theta _{0})}}
  {1+\ee ^{-s+\ii \, (\theta -\theta _{0})}}\, \dd s\, .\nonumber
\end{eqnarray}
The second term on the RHS of (\ref{eq:FinalGreenFce}) can be given
still another form with the aid of the identity
\begin{eqnarray*}
  &  & \int _{-\infty }^{\infty }K_{\ii \tau }(a)\, K_{-\ii \tau }(b)\,
  \frac{\ee ^{\phi \, \tau }}{\sin (\pi (\alpha +\ii \tau ))}\dd \tau \\
  &  & \qquad \quad =\, \int _{-\infty }^{\infty }K_{0}\!
  \left(\sqrt{a^{2}+b^{2}+2\, a\, b\, \cosh (u)}\, \right)
  \frac{\ee ^{-\alpha (u-\ii \, \phi )}}{1+\ee ^{-u+\ii \, \phi }}\, \dd u
\end{eqnarray*}
for $a>0$, $b>0$, $0<\alpha <1$ and $\left|\phi \right|<\pi $.
This way we get
\begin{subequations}\label{eq:FinalGreenFceSepar}
  \begin{eqnarray}
    \GG _{z}(r,\theta ;r_{0},\theta _{0})
    & = & \frac{1}{2\pi }K_{0}\!
    \left(\sqrt{-z}\left|x-x_{0}\right|\right)
    \label{eq:FinalGreenFceSepara}\\
    &  & -\frac{\sin (\pi \, \alpha )}{2\pi ^{2}}
    \int _{-\infty }^{\infty }K_{\ii \tau }\!
    \left(\sqrt{-z}\, r\right)\, K_{-\ii \tau }\!
    \left(\sqrt{-z}\, r_{0}\right)\,
    \frac{\ee ^{(\theta -\theta _{0})\, \tau }}
    {\sin (\pi (\alpha +\ii \, \tau ))}\, \dd \tau \, \nonumber
  \end{eqnarray}
  for $\theta -\theta _{0}\in (-\pi ,\pi )$,
  \begin{eqnarray}
    \GG _{z}(r,\theta ;r_{0},\theta _{0})
    & = & \ee ^{2\pi \, \ii \, \alpha }\Bigg (\frac{1}{2\pi }K_{0}
    \left(\sqrt{-z}\left|x-x_{0}\right|\right)
    \label{eq:FinalGreenFceSeparb}\\
    &  & -\frac{\sin (\pi \, \alpha )}{2\pi ^{2}}
    \int _{-\infty }^{\infty }K_{\ii \tau }\! \left(\sqrt{-z}\,r\right)\,
    K_{-\ii \tau }\! \left(\sqrt{-z}\, r_{0}\right)\,
    \frac{\ee ^{(\theta -\theta _{0}-2\pi )\, \tau }}
    {\sin (\pi (\alpha +\ii \, \tau ))}\, \dd \tau \Bigg )\nonumber
  \end{eqnarray}
  for $\theta -\theta _{0}\in (\pi ,2\pi )$, and
  \begin{eqnarray}
    \GG _{z}(r,\theta ;r_{0},\theta _{0})
    & = & \ee ^{-2\pi \, \ii \, \alpha }\Bigg (\frac{1}{2\pi }K_{0}\!
    \left(\sqrt{-z}\left|x-x_{0}\right|\right)
    \label{eq:FinalGreenFceSeparc}\\
    &  & -\frac{\sin (\pi \, \alpha )}{2\pi ^{2}}
    \int _{-\infty }^{\infty }K_{\ii \tau }\!
    \left(\sqrt{-z}\, r\right)\, K_{-\ii \tau }\!
    \left(\sqrt{-z}\, r_{0}\right)\,
    \frac{\ee ^{(\theta -\theta _{0}+2\pi )\, \tau }}
    {\sin (\pi (\alpha +\ii \, \tau ))}\, \dd \tau \Bigg )\nonumber
  \end{eqnarray}
  for $\theta -\theta _{0}\in (-2\pi ,-\pi )$.
\end{subequations}

Despite of this threefold description depending on the value of
$\theta -\theta_{0}$ the Green function should be continuous, even
real analytic, in its domain of definition if $x\neq x_{0}$. Checking
the limits from the right and left for $\theta -\theta _{0}=\pm \pi $
one finds that the continuity is guaranteed by the identity
\[
\int_{-\infty }^{\infty }K_{\ii\,\tau}(a)\, K_{-\ii\,\tau}(b)\dd\tau
= \pi \, K_{0}(a+b)\quad \textrm{for }\, a>0,\, b>0.
\]

Let us add a remark on deficiency subspaces. First we recall a general
and easy to verify fact. Let $A$ be a selfadjoint extension of a
symmetric operator $X$. Denote by $\NN (z)=\Ker (X^{*}-z)$ the deficiency
subspaces, $\Im\,z\neq 0$. Then it holds

\[
f\in \NN (w)\, \Longrightarrow \,
f+(z-w)(A-z)^{-1}f\in \NN(z)\,.
\]

This can be illustrated on our problem. We choose $H$ (the one-vortex
AB Hamiltonian defined by the boundary conditions
(\ref{eq:BCOneVortex})) for $A$, and $X$ is a restriction of $H$
obtained by requiring the supports of functions from the domain of $X$
to be separated from the singular point (the origin). The deficiency
indices are known to be $(2,2)$. For a basis in $\NN (z)$ we can
choose the vectors
\begin{equation}
\psi _{-1,z}(r,\theta )=K_{1-\alpha }\!
  \left(\sqrt{-z}\, r\right)\ee ^{\ii (\alpha -1)\theta },\textrm{
  }\psi _{0,z}(r,\theta )=K_{\alpha }\! \left(\sqrt{-z}\, r\right)\ee
  ^{\ii \, \alpha \, \theta }\,.
  \label{eq:DeficiencyBasisOneVort}
\end{equation}
Here $z\in \mathbb{C}\setminus \mathbb{R}_{+}$, $\Re \, \sqrt{-z}>0$.
As shown in Ref.~\citen{PSLDabrow} it holds true that
\begin{eqnarray*}
 &  & \int _{0}^{\infty }\left(\int _{-\pi }^{\pi }
   \GG _{z}(r,\theta ;r_{0},\theta _{0})\,
   \psi _{0,w}(r_{0},\theta _{0})\dd \theta _{0}\right)r_{0}\dd r_{0}\\
 &  & \qquad \quad =\, \frac{1}{z-w}\,\ee^{\ii \, \alpha \, \theta }
 \left(\left(\frac{\sqrt{-z}}{\sqrt{-w}}\right)^{\! \alpha }
   K_{\alpha }\! \left(\sqrt{-z}\, r\right)
   -K_{\alpha }\! \left(\sqrt{-w}\, r\right)\right),
\end{eqnarray*}
hence
\begin{equation}
\psi _{0,w}+(z-w)(H-z)^{-1}\psi
_{0,w}=\left(\frac{\sqrt{-z}}{\sqrt{-w}}\right)^{\! \alpha }\psi
_{0,z}\, .\label{eq:ImagePsi0}
\end{equation}
Similarly,

\begin{equation}
\psi _{-1,w}+(z-w)(H-z)^{-1}\psi _{-1,w}=\left(\frac{\sqrt{-z}}{\sqrt{-w}}\right)^{\! 1-\alpha }\psi _{-1,z}\, .\label{eq:ImagePsi-1}\end{equation}

Let us now focus on the case of two vortices but still considering
a spineless particle. The vortices are supposed to be located in the
points $a=(0,0)$ and $b=(\rho,0)$, $\rho >0$. Let $(r_{a},\theta _{a})$
be the polar coordinates centered at the point $a$ and $(r_{b},\theta _{b})$
be the polar coordinates centered at the point $b$. The two-vortex
AB Hamiltonian $H_{0}$ is the unique self-adjoint operator associated
to the quadratic form
\begin{subequations}\label{eq:QuadratFormTwoVort}
  \begin{equation}
    \int _{\mathbb{R}^{2}}\left(\left|\left(-\ii \, \partial_{x_{1}}
          -A_{1}\right)\! \varphi \right|^{2}+\left|\left(-\ii \,
          \partial _{x_{2}}-A_{2}\right)\! \varphi \right|\right)^{2}
    \dd^{2}x\,,\label{eq:QuadratFormTwoVorta}
  \end{equation}
  where
  \begin{equation}
    A=-\alpha \dd \theta _{a}-\beta \dd \theta_{b}.
    \label{eq:QuadratFormTwoVortb}
  \end{equation}
\end{subequations}Again, owing to the gauge equivalence, we can assume
that $\alpha ,\beta\in(0,1)$.

\begin{figure}[hp]
  {\centering \hspace*{-5mm}\includegraphics[scale=0.55]{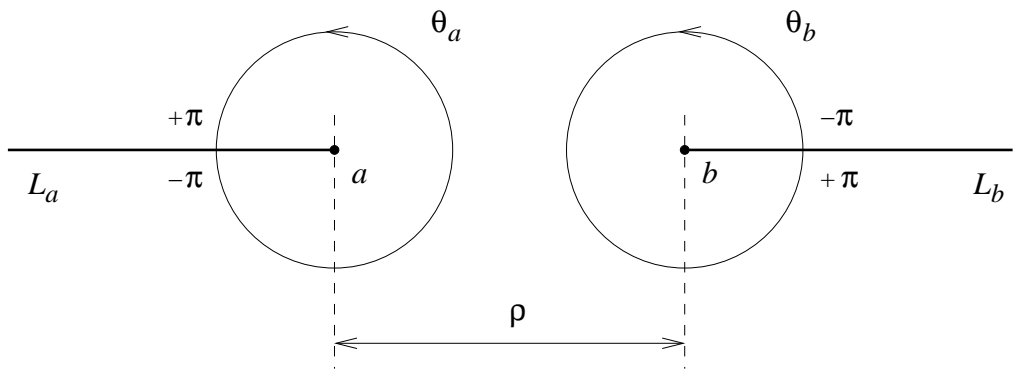}
    \par} \vskip 116pt
  \caption{Geometrical arrangement. Choice of the cuts $L_a$, $L_b$
    and choice of the angle variables $\theta_a$, $\theta_b$.}
  \label{fig:geom}
\end{figure}

Also in this case one can pass to a unitarily equivalent formulation.
The plane is cut along two half-lines,
\[
L_{a}=(-\infty ,0)\times \{0\}\textrm{ and }L_{b}=(\rho ,+\infty
)\times \{0\}.
\]
The values $\theta _{a}=\pm \pi$ correspond to the two sides of the
cut $L_{a}$ and similarly for $\theta _{b}$ and $L_{b}$.  The
geometrical arrangement is sketched in Fig.~\ref{fig:geom}. The
unitarily equivalent Hamiltonian $H$ is formally equal to $-\Delta$
and is determined by the boundary conditions along the cut,
\begin{eqnarray}
  &  & \hspace {-3em}\psi (r_{a},\theta _{a}=\pi )
  =\ee ^{2\pi \, \ii \, \alpha }\psi (r_{a},\theta _{a}
  =-\pi ),\textrm{ }\partial _{r_{a}}\psi (r_{a},\theta _{a}=\pi )
  =\ee ^{2\pi \, \ii \, \alpha }\partial _{r_{a}}\psi (r,\theta _{a}
  =-\pi )\, ,\nonumber \\
  & & \hspace {-3em}\psi (r_{b},\theta _{b}=\pi )=\ee ^{2\pi \, \ii \,
    \beta }\psi (r_{b},\theta _{b}=-\pi ),\textrm{ }\partial_{r_{b}}
  \psi (r_{b},\theta _{b}=\pi )=\ee ^{2\pi \, \ii \, \beta}
  \partial _{r_{b}}\psi (r,\theta _{b}=-\pi )\,.
  \label{eq:BoundaryCondsTwoVort}
\end{eqnarray}
In addition, one should impose a boundary condition at the vortex,
namely $\psi(a)=\psi(b)=0$.

A formula for the Green function of the Hamiltonian $H$ is known
also in the case of two vortices \cite{PS:PLA}. For a couple of points
$x,x_{0}\in \mathbb{R}^{2}\setminus(L_a\cup L_b)$ we set
\[
\zeta _{a}=1\textrm{ or }\zeta _{a}=\ee ^{2\, \pi \, \ii \, \alpha}
\textrm{ or }\zeta _{a}=\ee ^{-2\, \pi \, \ii \, \alpha }
\]
depending on whether the segment $\overline{x_{0}x}$ does not intersect
$L_{a}$, or $\overline{x_{0}x}$ intersects $L_{a}$ and $x_{0}$
lies in the lower half-plane, or $\overline{x_{0}x}$ intersects $L_{a}$
and $x_{0}$ lies in the upper half-plane. Analogously,
\[
\zeta _{b}=1\textrm{ or }\zeta _{b}=\ee ^{2\, \pi \, \ii \, \beta}
\textrm{ or }\zeta _{b}=\ee ^{-2\, \pi \, \ii \, \beta }
\]
depending on whether the segment $\overline{x_{0}x}$ does not intersect
$L_{b}$, or $\overline{x_{0}x}$ intersects $L_{b}$ and $x_{0}$
lies in the upper half-plane, or $\overline{x_{0}x}$ intersects $L_{b}$
and $x_{0}$ lies in the lower half-plane. Furthermore, let us set
\[
\zeta _{a}=\ee ^{\ii \, \alpha \, \eta _{a}},\, \, \zeta _{b}
=\ee^{\ii \, \beta \, \eta _{b}}\quad \textrm{where }
\eta _{a},\eta_{b}\in \{0,2\, \pi ,-2\, \pi \}.
\]

\begin{rem*}

Notice that if $\zeta _{a}\neq 1$ then necessarily $\zeta _{b}=1$
and vice versa.

\end{rem*}

The formula for the Green function reads
\begin{eqnarray}
  &  & \hspace {-2em}\GG _{z}(x,x_{0})
  =\zeta _{a}\zeta _{b}\frac{1}{2\, \pi }\, K_{0}\!
  \left(\sqrt{-z}\left|x-x_{0}\right|\right)\nonumber \\
  &  & -\, \zeta _{a}\frac{\sin (\pi \, \alpha )}{2\pi ^{2}}
  \int _{-\infty }^{\infty }K_{\ii \tau }\!
  \left(\sqrt{-z}\, r_{a}\right)\, K_{-\ii \tau }\!
  \left(\sqrt{-z}\, r_{0a}\right)\, \frac{\ee ^{(\theta _{a}
      -\theta _{0a}-\eta _{a})\tau }}{\sin (\pi (\alpha +\ii \tau ))}
  \dd \tau \nonumber \\
  &  & -\, \zeta _{b}\frac{\sin (\pi \, \beta )}{2\pi ^{2}}
  \int _{-\infty }^{\infty }K_{\ii \tau }\!
  \left(\sqrt{-z}\, r_{b}\right)\, K_{-\ii \tau }\!
  \left(\sqrt{-z}\, r_{0b}\right)\, \frac{\ee ^{(\theta _{b}
      -\theta _{0b}-\eta _{b})\tau }}{\sin (\pi (\beta +\ii \tau ))}
  \dd \tau \nonumber \\
  &  & +\, \frac{1}{2\pi }\sum _{\gamma ,\, n\geq 2}(-1)^{n}
  \int _{\mathbb{R}^{n}}K_{\ii \tau _{n}}\!
  \left(\sqrt{-z}\, r\right)K_{\ii (\tau _{n-1}-\tau _{n})}\!
  \left(\sqrt{-z}\, \rho \right)
  \label{eq:GreenFceTwoVort}\\
  &  & \quad \times \ldots \times K_{\ii (\tau _{1}-\tau _{2})}\!
  \left(\sqrt{-z}\, \rho \right)K_{-\ii \tau _{1}}\!
  \left(\sqrt{-z\, }r_{0}\right)\,
  \frac{\sin (\pi \, \sigma _{n})\, \exp (\theta \, \tau _{n})}
  {\pi \, \sin (\pi (\sigma _{n}+\ii \tau _{n}))}\nonumber \\
  &  & \quad \times \, \frac{\sin (\pi \, \sigma _{n-1})}
  {\pi \sin (\pi (\sigma _{n-1}+\ii \tau _{n-1}))}
  \times \ldots \times \frac{\sin (\pi \, \sigma _{2})}
  {\pi \sin (\pi (\sigma _{2}+\ii \tau _{2}))}\,
  \frac{\sin (\pi \, \sigma _{1})\, \exp (-\theta _{0}\tau _{1})}
  {\pi \sin (\pi (\sigma _{1}+\ii \tau _{1}))}\, \dd ^{n}\tau \, .\nonumber
\end{eqnarray}
Here the sum $\sum_{\gamma ,\, n\geq 2}$ runs over all finite
alternating sequences of length at least two, $\gamma
=(c_{n},c_{n-1},\ldots ,c_{1})$, such that for all $j$,
$c_{j}\in\{a,b\}$ and $c_{j}\neq c_{j+1}$, and $\sigma _{j}=\alpha $
(resp.  $\beta $) depending on whether $c_{j}=a$ (resp. $b$). In
addition, $(r,\theta )$ are the polar coordinates of the point $x$
with respect to the center $c_{n}$, $(r_{0},\theta _{0})$ are the
polar coordinates of the point $x_{0}$ with respect to the center
$c_{1}$(the dependence on $\gamma $ is not indicated explicitly).

\section{The Pauli Hamiltonian with AB
  fluxes\label{sec:The-Pauli-Hamiltonian}}

According to the Aharonov-Casher ansatz \cite{AharonovCasher} the two
diagonal components of the Pauli Hamiltonian with the third component
of spin equal to $\pm 1/2$ can be factorized,
\[
H^{\pm }=(p-A)^{2}\mp B=P_{\pm }^{*}P_{\pm }\,,
\]
where
\[
P_{\pm }=(p_{1}-A_{1})\pm \ii \, (p_{2}-A_{2}).
\]
Using the complex coordinate $z=x_{1}+\ii \, x_{2}$ one can rewrite
the Pauli Hamiltonian as follows:
\begin{eqnarray*}
  H^{+} & = & 4(-\ii \, \partial _{z}-A_{z})
  (-\ii \, \partial _{\overline{z}}-A_{\overline{z}}),\\
  H^{-} & = & 4(-\ii \, \partial _{\overline{z}}
  -A_{\overline{z}})(-\ii \, \partial _{z}-A_{z}).
\end{eqnarray*}

We start our discussion from considering the situation with one vortex.
Then
\[
A=\frac{\alpha }{r^{2}}(x_{2}\, \dd x_{1}-x_{1}\dd x_{2})=-\alpha \,
\dd \theta =\frac{\ii \, \alpha }{2}\, \ee ^{-2\ii \, \theta }
\dd \ee^{2\ii \, \theta }=\frac{\ii \, \alpha }{2}
\frac{\overline{z}}{z}\dd\frac{z}{\overline{z}}
=\frac{\ii \, \alpha }{2}\Big (\frac{\dd z}{z}
-\frac{\dd \overline{z}}{\overline{z}}\Big ).
\]
Hence
\[
A_{z}=\frac{\ii \, \alpha }{2z},\textrm{ }A_{\overline{z}}
=-\frac{\ii\, \alpha }{2\overline{z}},
\]
and we can write
\[
H^{+}=-4\Big (\, \partial _{z}+\frac{\alpha }{2z}\Big )
\Big (\,\partial _{\overline{z}}-\frac{\alpha }{2\overline{z}}\Big ),
\textrm{ }H^{-}=-4\Big (\, \partial_{\overline{z}}
-\frac{\alpha}{2\overline{z}}\Big )
\Big (\, \partial _{z}+\frac{\alpha }{2z}\Big ).
\]
In fact, these are formal expressions. More precisely, the operators
are defined as the unique selfadjoint operators associated
respectively to the positive quadratic forms

\begin{equation}
  \gq _{+}(\varphi )=4\int _{\mathbb{R}^{2}}\left|
    \Big (\partial_{\overline{z}}-\frac{\alpha }{2\overline{z}}\Big )
    \varphi\right|^{2}\dd ^{2}x\, ,\textrm{ }\gq _{-}(\varphi )
  =4\int_{\mathbb{R}^{2}}\left|\Big (\partial _{z}
    +\frac{\alpha }{2z}\Big)\varphi \right|^{2}\dd ^{2}x\, ,
  \label{eq:QFormsQpm}
\end{equation}
with their natural maximal domains of definition.

Since the magnetic field vanishes on $\mathbb{R}^{2}\setminus \{0\}$
the operators $H^{\pm }$ coincide with the spinless AB Hamiltonian
$H_{0}$ on the domain $\DD (\mathbb{R}^{2}\setminus \{0\})$ ($\DD $ is
the space of test functions). This means that all three operators
$H^{+}$, $H^{-}$ and $H_{0}$ are selfadjoint extensions of the same
symmetric operator $\widetilde{X}$. From Refs.~\citen{PSLDabrow} and
\citen{AdamiTeta} it is known that all selfadjoint extensions can be
described by appropriate boundary conditions at the origin. The method
used to derive the boundary conditions was inspired by the description
of point interactions in the plane given in
Ref.~\citen{Albeverioetal}.  Let us also note that analogous boundary
conditions have been derived in Ref.~\citen{PS:withEV} for the model
with additional homogeneous magnetic field while the case of
Dirac-Weyl operator is discussed in Ref.~\citen{Ogurisu}.

To describe the boundary conditions one introduces four functionals,
\begin{eqnarray*}
  \Phi _{1}^{-1}(\varphi ) & = & \lim _{r\downarrow 0}\, r^{1-\alpha }
  \frac{1}{2\pi }\int _{0}^{2\pi }\varphi (r,\theta )\,
  \ee ^{\ii \, \theta }\dd \theta ,\\
  \Phi _{2}^{-1}(\varphi ) & = & \lim _{r\downarrow 0}\, r^{-1+\alpha }
  \left(\frac{1}{2\pi }\int _{0}^{2\pi }\varphi (r,\theta )\,
    \ee ^{\ii \, \theta }\dd \theta -r^{-1+\alpha }
    \Phi _{1}^{-1}(\varphi )\right),\\
  \Phi _{1}^{0}(\varphi ) & = & \lim _{r\downarrow 0}\, r^{\alpha }
  \frac{1}{2\pi }\int _{0}^{2\pi }\varphi (r,\theta )\dd \theta ,\\
  \Phi _{2}^{0}(\varphi ) & = & \lim _{r\downarrow 0}\, r^{-\alpha }
  \left(\frac{1}{2\pi }\int _{0}^{2\pi }\varphi (r,\theta )\dd \theta
    -r^{-\alpha }\Phi _{1}^{0}(\varphi )\right).
\end{eqnarray*}
Each boundary condition is determined by a couple of matrices
$A_{1},A_{2}\in \Mat (2,\mathbb{C})$ fulfilling (the symbol
$(A_{1},A_{2})$ designates a $2\times4$ matrix)
\[
\rank\,(A_{1},A_{2})=2,\textrm{ }
A_{1}D^{-1}A_{2}^{*}=A_{2}D^{-1}A_{1}^{*}
\]
where
\[
D=
\begin{pmatrix}
  1-\alpha  & 0\\
  0 & \alpha
\end{pmatrix}.
\]
The boundary condition takes the form
\[
A_{1}
\begin{pmatrix}
  \Phi _{1}^{-1}(\varphi )\\
  \Phi _{1}^{0}(\varphi )
\end{pmatrix}
+ A_{2}
\begin{pmatrix}
  \Phi _{2}^{-1}(\varphi )\\
  \Phi _{2}^{0}(\varphi )
\end{pmatrix} =
\begin{pmatrix}
  0\\
  0
\end{pmatrix}.
\]
Two couples of matrices, $\{A_{1},A_{2}\}$ and $\{A'_{1},A'_{2}\}$,
determine the same boundary condition if and only if there exists a
regular matrix $G\in \GL (2,\mathbb{C})$ such that
$(A'_{1},A'_{2})=G(A_{1},A_{2})$.

For example, the domain of the spinless AB Hamiltonian $H_{0}$ is
determined by the boundary conditions at the vortex
$\Phi_{1}^{-1}(\varphi )=\Phi _{1}^{0}(\varphi )=0$ and so by the couple of
matrices
\[
A_{1}=
\begin{pmatrix}
  1 & 0\\
  0 & 1
\end{pmatrix},
\textrm{ }A_{2}=
\begin{pmatrix}
  0 & 0\\
  0 & 0
\end{pmatrix}.
\]

We wish to derive the boundary conditions for the Hamiltonians $H^{+}$
and $H^{-}$. According to the well-known construction, the operator
$A$ associated to a semi-bounded quadratic form $\gq $ is determined
by the condition
\[
\forall f\in\Dom(A)\subset\Dom(\gq),\,
\forall\varphi\in\Dom(\gq ),\textrm{ }\langle \varphi, Af\rangle
=\gq (\varphi ,f)\, .
\]
This is to say that $f\in \Dom (\gq )$ belongs to $\Dom (A)$ if and
only if there exists $g\in \HH $ such that the equality
$\langle\varphi,g\rangle =\gq (\varphi ,f)$ holds true for all
$\varphi\in\Dom(\gq )$. In that case $g$ is unique and $Af=g$. We are
going to apply this prescription to the quadratic forms
(\ref{eq:QFormsQpm}).  This amounts to integration by parts.

More precisely, the Green formula implies that
\[
\int _{\mathbb{R}^{2}}(\partial _{z}f)g\dd ^{2}x
=-\int_{\mathbb{R}^{2}}f(\partial _{z}g)\dd ^{2}x
-\lim _{a\downarrow 0}\,
\frac{a}{2}\int _{0}^{2\pi }(f\, g)(a\, \cos (\theta ),a\,
\sin(\theta ))\, \ee ^{-\ii \, \theta }\dd \theta \, .
\]
Thus one finds that $f\in\Dom(\widetilde{X}^{*})$ belongs to
$\Dom(H^{+})$ if and only if for all $\varphi \in \Dom (\gq _{+})$,
\[
\lim _{a\downarrow 0}\, a\int _{0}^{2\pi }\Big (\overline{\varphi }\,
\Big (\, \partial _{\overline{z}}-\frac{\alpha }{2\overline{z}}\Big
)f\Big )(a\, \cos (\theta ),a\, \sin (\theta ))\, {{\ee }^{-\ii \,
    \theta }}\dd \theta \, =0,
\]
or, when expressing $(z,\overline{z})$ in the polar coordinates,
\[
\lim _{a\downarrow 0}\, \int _{0}^{2\pi }\big (\overline{\varphi }\,
(\, r\, \partial _{r}+\ii \, \partial _{\theta }-\alpha )f\big )
(a\cos (\theta ),a\sin (\theta ))\dd \theta \, =0.
\]

Any $f\in\Dom(\widetilde{X}^{*})$ asymptotically behaves like
\[
f=\left(\Phi _{1}^{-1}(f)\, r^{-1+\alpha }+\Phi _{2}^{-1}(f)\,
  r^{1-\alpha }\right)\ee ^{-\ii \, \theta }+\left(\Phi _{1}^{0}(f)\,
  r^{-\alpha }+\Phi _{2}^{0}(f)\, r^{\alpha }\right)
+\textrm{ regular part}.
\]
Hence
\[
(r\, \partial _{r}+\ii \, \partial _{\theta }-\alpha )f
\sim 2(1-\alpha)\Phi _{2}^{-1}(f)\, r^{1-\alpha }
\ee ^{-\ii \, \theta }-2\alpha \,
\Phi _{1}^{0}(f)\, r^{-\alpha }+\ldots \, .
\]
Notice that
\[
(r\, \partial _{r}+\ii \, \partial _{\theta }-\alpha )r^{-1+\alpha }
\ee ^{-\ii \, \theta }=(r\, \partial _{r}+\ii \, \partial _{\theta}
-\alpha )r^{\alpha }=0
\]
and so any function of the form
$r^{-1+\alpha}\eta(r)\ee^{-\ii\,\theta}$ or $r^{\alpha}\eta(r)$, with
$\eta\in{}C^{\infty}(\mathbb{R}_{+})$, $\eta(r)\equiv1$ in a
neighborhood of 0 and $\eta(r)\equiv 0$ in a neighborhood of
$+\infty$, belongs to $\Dom(\gq _{+})$.  Therefore a sufficient and
necessary condition for $f$ to belong to $\Dom (H^{+})$ is
\begin{equation}
  \Phi _{2}^{-1}(f)\, =\Phi_{1}^{0}(f)=0.
  \label{eq:BoundCondHplus}
\end{equation}
The corresponding couple of matrices can be chosen as

\[
A_{1}=
\begin{pmatrix}
  0 & 1\\
  0 & 0
\end{pmatrix},
\textrm{ }A_{2}=
\begin{pmatrix}
  0 & 0\\
  1 & 0
\end{pmatrix}
\]

The other component of the Pauli Hamiltonian, $H^{-}$, can be treated
similarly. One finds that $f\in\Dom(\widetilde{X}^{*})$ belongs to
$\Dom(H^{-})$ if and only if for all $\varphi\in\Dom(\gq _{-})$,
\[
\lim _{a\downarrow 0}\, \int _{0}^{2\pi }\big (\overline{\varphi }\,
(r\,\partial _{r}-\ii \, \partial _{\theta }+\alpha )f\big )
(a\cos (\theta),a\sin (\theta ))\dd \theta =0
\]
which turns out to be equivalent to
\begin{equation}
  \Phi _{1}^{-1}(f)=\Phi _{2}^{0}(f)=0.
  \label{eq:BoundCondHminus}
\end{equation}
The corresponding couple of matrices can be chosen as

\[
A_{1}=
\begin{pmatrix}
  1 & 0\\
  0 & 0
\end{pmatrix},\textrm{ }A_{2}=
\begin{pmatrix}
  0 & 0\\
  0 & 1
\end{pmatrix}.
\]

The generalization to the case of several vortices is quite straightforward.
One simply imposes the above derived boundary conditions at each vortex.
Let us consider the case of two vortices. For the sake of simplicity
we assume that the vortices are $a=(0,0)$ and $b=(1,0)$. The Pauli
Hamiltonian formally reads
\begin{eqnarray*}
  H^{+} & = & -4\left(\partial _{z}+\frac{1}{2}\Big (\frac{\alpha }{z}
    +\frac{\beta }{z-1}\Big )\right)\left(\partial _{\overline{z}}
    -\frac{1}{2}\Big (\frac{\alpha }{\overline{z}}
    +\frac{\beta }{\overline{z}-1}\Big )\right),\\
  H^{-} & = & -4\left(\partial _{\overline{z}}-\frac{1}{2}
    \Big (\frac{\alpha }{\overline{z}}+\frac{\beta }
    {\overline{z}-1}\Big )\right)\left(\partial _{z}+\frac{1}{2}
    \Big (\frac{\alpha }{z}+\frac{\beta }{z-1}\Big )\right).
\end{eqnarray*}
We still assume that $0<\alpha ,\beta <1$ (in virtue of the gauge
equivalence).

The Pauli Hamiltonian with two vortices is known to have zero modes
\cite{Arai}. They can be computed with the aid of the Aharonov-Casher
ansatz since it effectively enables to reduce the second order
differential equation to a first order one. Explicit solutions are
even known in some essentially more complicated situations (see for
example Ref.~\citen{GeylerGrishanov}). Just for the sake of
illustration let us verify that the zero modes actually satisfy the
above derived boundary conditions (\ref{eq:BoundCondHplus}) or
(\ref{eq:BoundCondHminus}).

If $\alpha +\beta <1$ then the function
\[
\varphi (z)=\frac{\left|z\right|^{\alpha }\left|z-1\right|^{\beta}}{z(z-1)}
\]
is $L^{2}$ integrable and solves
\[
\left(\partial _{\overline{z}}-\frac{1}{2}\Big (\frac{\alpha}
  {\overline{z}}+\frac{\beta }{\overline{z}-1}\Big )\right)\varphi =0.
\]
So it is a zero mode of $H^{+}$. It is elementary to compute its
asymptotic behavior for $r_{a}\rightarrow 0$,
\[
\varphi =r_{a}^{-1+\alpha }\ee ^{-\ii \, \theta _{a}}
+\Big (1-\frac{\beta}{2}\Big )r_{a}^{\alpha }
-\frac{\beta }{2}\, r_{a}^{\alpha }\ee ^{-2\,
  \ii \, \theta _{a}}+O\big (r_{a}^{1+\alpha }\big ).
\]
Hence $\varphi $ obeys (\ref{eq:BoundCondHplus}). The boundary condition
at the vortex $b$ is analogous.

Similarly, if $\alpha +\beta >1$ then
\[
\varphi (z)=\frac{1}{\left|z\right|^{\alpha }\left|z-1\right|^{\beta }}
\]
is a zero mode of $H^{-}$ and
\[
\varphi =-\frac{\beta }{2}\, r_{a}^{1-\alpha }
\ee ^{-\ii \, \theta_{a}}+r_{a}^{-\alpha }
-\frac{\beta }{2}\, r_{a}^{1-\alpha }\ee ^{\ii
  \, \theta _{a}}+O\big (r_{a}^{2-\alpha }\big ).
\]
Hence $\varphi$ obeys (\ref{eq:BoundCondHminus}).

\section{Asymptotic behavior near a vortex\label{sec:Asymptotic-behaviour}}

Our first task in this section is the asymptotic analysis of functions
from a deficiency subspace. To simplify the discussion we shall use
the symbol $O(r^{\gamma })$ in a sense somewhat weaker than it is
common. The equality $f(r,\theta )=O(r^{\gamma })$ for $r\downarrow 0$
will mean that
$f(r,\theta)=\sum_{n\in\mathbb{Z}}f_{n}(r)\ee^{\ii{}n\theta }$ and for
all $n$ it holds $f_{n}(r)=O(r^{\gamma })$.

\begin{lem}\label{thm:PsiOnDiskGenSing}

  Assume that $R>0$, $z\in\mathbb{C}\setminus\mathbb{R}_{+}$,
  $0<\alpha <1$ and $\varphi\in L^{2}(B(0,R),\dd ^{2}x)$ satisfies in
  the weak sense the differential equation
  \[
  (Y-z)\varphi = 0
  \]
  on $B(0,R)\setminus \{0\}$ (the disk centered at $0$ with the radius
  equal to $R$) where (using the polar coordinates $(r,\theta)$)
  \begin{eqnarray*}
    Y & = & -\ee ^{-\ii \, \alpha \, \theta }\Delta \,
    \ee ^{\ii \, \alpha \, \theta }\, =\, -(\partial _{x_{1}}+
    \ii \, \alpha \, \partial _{x_{1}}\theta )^{2}-(\partial _{x_{2}}
    +\ii \, \alpha \, \partial _{x_{2}}\theta )^{2}\\
    & = & -\left(\frac{\partial ^{2}}{\partial
        r^{2}}+\frac{1}{r}\frac{\partial }{\partial
        r}+\frac{1}{r^{2}}\Big (\frac{\partial }{\partial \theta }+\ii
      \, \alpha \Big )^{2}\right).
  \end{eqnarray*}
  Then there exist constants $c_{0}$, $d_{0}$, $c_{-1}$, $d_{-1}$,
  such that
  \begin{equation}
    \varphi(r,\theta ) = c_{0}r^{-\alpha}+d_{0}r^{\alpha }
    +\big (c_{-1}r^{-1+\alpha }+d_{-1}r^{1-\alpha}\big )
    \ee ^{-\ii \, \theta }+O(r^{\gamma })\quad \textrm{for }
    r\downarrow 0\,,
    \label{eq:PsiOnDiskGenSing}
  \end{equation}
  where $\gamma =\min \{2-\alpha ,1+\alpha \}$.

\end{lem}

\begin{proof}

  For all $n\in\mathbb{Z}$, $\eta\in\DD((0,R))$ (the space of test
  functions) it holds true that
  \[
  0 = \langle (Y-\bar{z})\eta(r)\,
  \ee ^{\ii \, n\, \theta },\varphi \rangle
  = -\Big \langle\left(\frac{\partial ^{2}}{\partial r^{2}}
    +\frac{1}{r}\frac{\partial}{\partial r}
    + \bar{z} - \frac{(n+\alpha )^{2}}{r^{2}}\right)\! \eta(r)\,
  \ee ^{\ii \, n\, \theta },\varphi \Big \rangle .
  \]
  Hence
  \[
  \varphi(r,\theta )
  = \sum _{n=-\infty }^{\infty }f_{n}(r)\, \ee ^{\ii \, n\,
    \theta }\,,\] where\[ \forall n\in \mathbb{Z},\textrm{
  }\left(\frac{\partial ^{2}}{\partial
      r^{2}}+\frac{1}{r}\frac{\partial }{\partial r}+z-\frac{(n+\alpha
      )^{2}}{r^{2}}\right)\! f_{n}(r)=0\quad \textrm{on }(0,R)
  \]
  in the weak sense. This implies that the generalized derivative
  $\partial_r\big(r\,\partial_rf_n(r)\big)$ belongs to
  $L^1_{\mathrm{loc}}\big((0,R)\big)$ and consequently
  $f_n\in{}AC^2\big((\varepsilon,R)\big)$ for all $0<\varepsilon<R$.
  Therefore necessarily $f_{n}(r)$ is a linear combination of the
  modified Bessel functions,
  \[
  f_{n}(r)=a_{n}\, K_{n+\alpha }\! \left(\sqrt{-z}\,
    r\right)+b_{n}\, I_{\left|n+\alpha \right|}\! \left(\sqrt{-z}\,
    r\right).\] Let us recall the asymptotic behavior of the Bessel
  functions. If $0<\nu $ and $\nu \notin \mathbb{N}$ then\[ I_{\nu
  }(r)=\frac{1}{\Gamma (\nu +1)}\Big (\frac{r}{2}\Big )^{\nu }+O\big
  (r^{\nu +2}\big )
  \]
  and
  \[
  K_{\nu }(r)=\frac{\Gamma (\nu )}{2}\left(\frac{r}{2}\right)^{-\nu
  }\left(1+O\big (r^{2}\big )\right)-\frac{\Gamma (1-\nu )}{2\nu
  }\left(\frac{r}{2}\right)^{\nu }\left(1+O\big (r^{2}\big )\right).
  \]
  This implies that $f_{n}\in L^{2}((0,R),r\dd r)$ if and only if
  either $a_{n}=0$ or $n\in \{0,-1\}$. This is to say that $a_{n}$ can
  be nonzero only for $n=0,-1$. So if $n\neq 0,-1$ then
  $f_{n}(r)=O(r^{\left|n+\alpha \right|})$. This proves the lemma.
\end{proof}

Let $H$ be the two-vortex spinless AB Hamiltonian defined by boundary
conditions (\ref{eq:BoundaryCondsTwoVort}). The symbol $X$ below
stands for the symmetric operator obtained by restricting the domain
of $H$ so that functions from $\Dom X$ vanish in some neighborhood of
the vortices. The deficiency subspaces are denoted by
$\NN(z)=\Ker(X^{*}-z)$.

\begin{cor}\label{thm:PsiFromDeficiSpaceGenSing}

  If $z\in \mathbb{C}\setminus \mathbb{R}_{+}$ and $\psi\in\NN(z)$
  then there exist constants $c_{a,0}$, $c_{a,-1}$, $c_{b,0}$,
  $c_{b,-1}$, such that
  \begin{equation}
    \psi (x)=c_{a,0}\, r_{a}^{-\alpha }\, \ee ^{\ii \, \alpha \,
      \theta_{a}}+c_{a,-1}\, r_{a}^{-1+\alpha }\, \ee ^{\ii (\alpha -1)
      \theta_{a}}+o(1)\quad \textrm{for }\, r_{a}\downarrow 0,
    \label{eq:PsiFromDeficiSpaceGenSinga}
  \end{equation}
  and
  \begin{equation}
    \psi (x)=c_{b,0}\, r_{b}^{-\beta }\, \ee ^{\ii \, \beta \,
      \theta_{b}}+c_{b,-1}\, r_{b}^{-1+\beta }\, \ee ^{\ii (\beta -1)
      \theta_{b}}+o(1)\quad \textrm{for }\, r_{b}\downarrow 0.
    \label{eq:PsiFromDeficiSpaceGenSingb}
  \end{equation}

\end{cor}

\begin{proof}
  The property $\psi \in \NN (z)$ means that
  $\psi\in{}L^{2}(B(0,R),\dd ^{2}x)$, $(-\Delta -z)\psi =0$ on
  $\mathbb{R}^{2}\setminus (L_{a}\cup L_{b})$ in the weak sense and
  $\psi $ satisfies the boundary conditions
  (\ref{eq:BoundaryCondsTwoVort}) on $L_{a}\cup L_{b}$. Then the
  function $\exp(-\ii\,\alpha\,\theta _{a})\psi $ obeys the
  assumptions of Lemma~\ref{thm:PsiOnDiskGenSing} and relation
  (\ref{eq:PsiOnDiskGenSing}) implies
  (\ref{eq:PsiFromDeficiSpaceGenSinga}). Relation
  (\ref{eq:PsiFromDeficiSpaceGenSingb}) can be shown similarly.
\end{proof}

\begin{cor}\label{thm:PsiDomHVanish}

  Assume that $z\in\mathbb{C}\setminus\mathbb{R}_{+}$, $\psi\in\NN(z)$
  and $\psi(a)=\psi(b)=0$. Then $\psi\in\Dom(H)$ and hence $\psi=0$.

\end{cor}

\begin{proof}
  We use once more the fact that $\exp(-\ii\,\alpha\,\theta_{a})\psi$
  obeys the assumptions of Lemma~\ref{thm:PsiOnDiskGenSing} and hence
  \begin{displaymath}
    \exp(-\ii\,\alpha\,\theta_{a})\psi(x)
    = c_{0}r_a^{\,-\alpha}+d_{0}r_a^{\,\alpha }
    +\big (c_{-1}r_a^{\,-1+\alpha }+d_{-1}r_a^{\,1-\alpha}\big )
    \ee^{-\ii \, \theta_a}+O(r_a^{\,\gamma })\quad \textrm{for }
    r_a\downarrow 0\,.
  \end{displaymath}
  Since $\psi(a)=0$ it holds $c_{-1}=c_0=0$. Let $U$ be the unitary
  operator on $L^2(\mathbb{R}^2,\dd^2x)$ acting via multiplication by
  the phase factor $\exp(\ii\alpha\theta_{a}+\ii\beta\theta_{b})$.
  Then $\varphi=\exp(-\ii\alpha\theta_{a}-\ii\beta\theta_{b})\psi$
  belongs to $\Ker(\widetilde{X}^\ast-z)$ where
  $\widetilde{X}=U^{-1}XU$. The function $\theta_b(x)$ is real
  analytic in a neighborhood of $a$ and
  \begin{displaymath}
    \theta_b(x) = \sin(\theta_a)\,\frac{r_a}{\rho}+O(r_a^{\,2})
    \quad \textrm{for }r_a\downarrow 0\,.
  \end{displaymath}
  A straightforward computation gives the asymptotic behavior of
  $\varphi$ and one finds that
  \begin{displaymath}
    \Phi^{-1}_1(\varphi) = c_{-1},\textrm{ }
    \Phi^{-1}_2(\varphi) = d_{-1}+\frac{\beta c_0}{2\rho},\textrm{ }
    \Phi^0_1(\varphi) = c_0,\textrm{ }
    \Phi^0_2(\varphi) = d_0-\frac{\beta c_{-1}}{2\rho}\,.
  \end{displaymath}
  So one finds that the boundary condition
  $\Phi^{-1}_1(\varphi)=\Phi^0_1(\varphi)=0$ is satisfied at the
  vortex $a$. Analogously, the same boundary condition is fulfilled at
  the vortex $b$. As recalled in
  Section~\ref{sec:The-Pauli-Hamiltonian}, these boundary conditions
  determine the domain of $H_0$. Hence $\varphi\in\Dom(H_0)$ and
  $\psi\in\Dom(H)$. But $H$ is positive, $z\not\in\mathbb{R}_+$, and
  therefore $\Dom(H)\cap\NN(z)=\{0\}$. This shows that $\psi=0$.
\end{proof}

Further we are interested in the asymptotic behavior near a vortex of
the Green functions (\ref{eq:FinalGreenFceSepar}) and
(\ref{eq:GreenFceTwoVort}).  It is easy to see that in the spinless
case the Green function vanishes in each vortex. For example in the
case of two vortices it holds true that $\GG _{z}(a,x_{0})=0.$ This
can be derived from (\ref{eq:GreenFceTwoVort}) with the aid of the
relation
\begin{equation}
  K_{\ii \tau }(r)\rightarrow \pi \, \delta(\tau )\quad
  \textrm{for }r\downarrow 0
  \label{eq:LimitKtoDelta}
\end{equation}
and some simple combinatorics. It is also obvious that
\[
K_{0}\!\left(\sqrt{-z}\left|x-x_{0}\right|\right)
= K_{0}\! \left(\sqrt{-z}\,
  r_{0a}\right)+O(r_{a})\quad \textrm{for }r_{a}\downarrow 0,
\]
(here $r_{0a}=\left|a-x_{0}\right|$) and
\[
K_{\ii \nu }\! \left(\sqrt{-z}\,
  r_{b}\right)=K_{\ii \nu }\! \left(\sqrt{-z}\, \rho \right)
+O(r_{a})\quad \textrm{for }r_{a}\downarrow 0.
\]

To get an additional information we shall need an asymptotic formula
for the integral
\begin{equation}
  \int _{-\infty }^{\infty }K_{\ii \tau }\!
  \left(\sqrt{-z}\, r_{a}\right)\, K_{\ii (\nu -\tau )}\!
  \left(\sqrt{-z}\, \rho \right)\frac{\sin (\pi \alpha )\,
    \exp (\theta _{a}\, \tau )}{\pi \, \sin(\pi(\alpha+\ii\tau))}\,
  \dd \tau \, .
  \label{eq:IntKK}\end{equation}
Such an asymptotic analysis can be carried on with the aid of the
following lemma.

\begin{lem}\label{thm:AsymptExpChFracExp}
  \begin{subequations}\label{eq:AsymptExpChFracExp}

    Suppose that $r>0$, $|\theta |<\pi $ and $0<\alpha <1$. Then
    \begin{eqnarray}
      \int _{-\infty }^{\infty }\ee^{-r\cosh (s)}\,
      \frac{\ee ^{-\alpha (u-s-\ii \theta )}}
      {1+\ee ^{-u+s+\ii \theta }}\dd s
      & = & \frac{\pi }{\sin (\pi \, \alpha )}
      -\frac{\Gamma (\alpha )}{1-\alpha }
      \left(\frac{r}{2}\right)^{\! 1-\alpha }
      \ee ^{(1-\alpha )(u-\ii \theta )}\nonumber \\
      &  & -\, \frac{\Gamma (1-\alpha )}{\alpha }
      \left(\frac{r}{2}\right)^{\! \alpha }
      \ee ^{-\alpha (u-\ii \theta )}+Z(r,u)\,,
      \label{eq:AsymptExpChFracExpa}
    \end{eqnarray}
    where

    \begin{equation}
      \forall r\in (0,1),\, \forall u\in \mathbb{R},
      \textrm{ }\left|Z(r,u)\right|\leq Kr\cosh (u)
      \label{eq:AsymptExpChFracExpb}
    \end{equation}
    and $K$ depends on $\theta $ and $\alpha $ but does not depend on
    $r$ and $u$.
  \end{subequations}
\end{lem}

\begin{proof}

The LHS of (\ref{eq:AsymptExpChFracExpa}) equals
\begin{equation}
  \int _{u}^{\infty }\ee^{-r\cosh (s)}\,
  \frac{\ee ^{-(1-\alpha )(s-u+\ii\theta)}}
  {1+\ee^{-s+u-\ii \theta}}\dd s
  + \int _{-u}^{\infty }\ee^{-r\cosh (s)}\,
  \frac{\ee ^{-\alpha (s+u-\ii \theta )}}
  {1+\ee ^{-s-u+\ii\theta }}\dd s\, .
  \label{eq:AsymptExpInter1}
\end{equation}
Therefore it suffices to study integrals of the form
\begin{eqnarray}
  \int _{u}^{\infty }\ee ^{-r\cosh (s)}
  \frac{\ee ^{-\gamma (s-u+\ii \theta )}}{1+\ee^{-s+u-\ii\theta}}\dd s
  & = & \ee ^{\gamma (u-\ii \theta )}\int _{u}^{\infty }
  \ee ^{-r\cosh (s)-\gamma s}\dd s
  \nonumber \\
  &  & -\int _{u}^{\infty }\ee ^{-r\cosh (s)}\,
  \frac{\ee ^{-(\gamma +1)(s-u+\ii \theta )}}
  {1+\ee ^{-s+u-\ii \theta }}\dd s
  \label{eq:AsymptExpInter2}
\end{eqnarray}
for $0<\gamma <1$. The second integral on the RHS of
(\ref{eq:AsymptExpInter2}) can be treated easily and one finds that
\[
\int _{u}^{\infty }\ee ^{-r\cosh (s)}\,
\frac{\ee ^{-(\gamma+1)(s-u+\ii \theta )}}
{1+\ee ^{-s+u-\ii \theta }}\dd s
=\frac{\ee^{-\ii \gamma \theta }}{\gamma }
-\int _{0}^{\infty }\frac{\ee^{-\gamma (s+\ii \theta )}}
{1+\ee ^{-s-\ii \theta }}\dd s+Z_{1}(r,u)\,,
\]
where $Z_{1}(r,u)$ satisfies estimate (\ref{eq:AsymptExpChFracExpb}).
To treat the first integral on the RHS of (\ref{eq:AsymptExpInter2})
we note that\[
\ee ^{-r\cosh (s)}=\exp \! \left(-\frac{r}{2}\, \ee ^{s}\right)
\sum_{k=0}^{\infty }\frac{1}{k!}\left(-\frac{r}{2}\right)^{\! k}\ee^{-ks}
\]
and therefore
\begin{eqnarray*}
  \int _{0}^{\infty }\ee ^{-r\cosh (s)-\gamma s}\dd s
  & = & \sum _{k=0}^{\infty }\frac{1}{k!}\, 2^{-\gamma-k}r^{\gamma+k}\,
  \Gamma \! \left(-\gamma -k,\frac{r}{2}\right)
  \left(-\frac{r}{2}\right)^{\! k}\\
  & = & -\frac{\Gamma (1-\gamma )}{\gamma }
  \left(\frac{r}{2}\right)^{\! \gamma }+\frac{1}{\gamma }
  +\frac{\gamma }{1-\gamma ^{2}}\, r+O(r^{2})\, .
\end{eqnarray*}
Furthermore,
\[
\int _{u}^{0}\ee ^{-r\cosh (s)-\gamma s}\dd s
=\frac{\ee ^{-\gamma u}-1}{\gamma }+Z_{2}(r,u)\,,
\]
where $Z_{2}(r,u)$ satisfies estimate (\ref{eq:AsymptExpChFracExpb}).
Thus we have derived that
\begin{equation}
  \int _{u}^{\infty }\ee ^{-r\cosh (s)}
  \frac{\ee ^{-\gamma (s-u+\ii \theta )}}{1+\ee^{-s+u-\ii\theta}}\dd s
  =-\frac{\Gamma (1-\gamma )}{\gamma }
  \left(\frac{r}{2}\right)^{\! \gamma }\ee ^{\gamma (u-\ii \theta )}
  +\int _{0}^{\infty }\frac{\ee ^{-\gamma (s+\ii \theta )}}
  {1+\ee ^{-s-\ii \theta }}\dd s+Z_{3}(r,u)\,,
  \label{eq:AsymptExpInter3}
\end{equation}
where $Z_{3}(r,u)$ satisfies estimate (\ref{eq:AsymptExpChFracExpb}).
To conclude the proof it suffices to apply (\ref{eq:AsymptExpInter3})
to the both integrals in (\ref{eq:AsymptExpInter1}) and to take into
account that
\[
\int _{0}^{\infty }\frac{\ee ^{-(1-\alpha )(s+\ii \theta )}}
{1+\ee^{-s-\ii \theta }}\dd s+\int _{0}^{\infty }
\frac{\ee ^{-\alpha (s-\ii \theta )}}{1+\ee ^{-s+\ii \theta }}\dd s
=\int _{-\infty}^{\infty }\frac{\ee ^{-\alpha (s-\ii \theta )}}
{1+\ee ^{-s+\ii \theta }}\dd s=\frac{\pi }{\sin (\pi \alpha )}
\]
for $|\theta |<\pi $.\end{proof}

\begin{cor}

Under the same assumptions as in Lemma~\ref{thm:AsymptExpChFracExp}
it holds true that
\begin{eqnarray}
  &  & \hspace {-2em}\int _{-\infty }^{\infty }K_{\ii \tau }\!
  \left(\sqrt{-z}\, r\right)\, K_{\ii (\nu -\tau )}\!
  \left(\sqrt{-z}\, \rho \right)\frac{\sin (\pi \alpha )\,
    \exp (\theta \, \tau )}{\pi \, \sin (\pi (\alpha +\ii \tau ))}
  \dd \tau \nonumber \\
  &  & =\, K_{\ii \nu }\! \left(\sqrt{-z}\, \rho \right)
  -\frac{\sin (\pi \, \alpha )}{\pi }\frac{\Gamma (\alpha )}{1-\alpha}
  \left(\frac{\sqrt{-z}\, r}{2}\right)^{\! 1-\alpha }
  \ee ^{\ii (\alpha -1)\theta }K_{\ii \nu -1+\alpha }\!
  \left(\sqrt{-z}\, \rho \right)
  \nonumber \\
  &  & \quad -\, \frac{\sin (\pi \, \alpha )}{\pi }
  \frac{\Gamma (1-\alpha )}{\alpha }\left(
    \frac{\sqrt{-z}\, r}{2}\right)^{\! \alpha }
  \ee ^{\ii \, \alpha \, \theta }K_{\ii \nu +\alpha }\!
  \left(\sqrt{-z}\, \rho \right)+O(r)
  \label{eq:AsymptIntKK}
\end{eqnarray}
for $r\downarrow 0$.

\end{cor}

\begin{proof}
  Using
  \begin{equation}
    K_{\ii \tau }(a)=\frac{1}{2}\int _{-\infty }^{\infty }
    \ee ^{\ii \, s\, \tau -a\cosh (s)}\dd s\quad
    \textrm{for }a>0,\, \tau \in \mathbb{R},
    \label{eq:KimIntegralForm}
  \end{equation}
  and applying the equality
  \[
  \int _{-\infty }^{\infty }\ee ^{\ii \, \tau (u-s)}
  \frac{\exp (\theta\, \tau )}{\sin (\pi (\alpha +\ii \tau ))}
  \dd \tau =2\, \frac{\ee^{-\alpha (u-s-\ii \theta )}}
  {1+\ee ^{-u+s+\ii \theta }}
  \]
  we find that (\ref{eq:IntKK}) equals
  \[
  \frac{\sin (\pi \, \alpha )}{2\pi }\int _{-\infty }^{\infty }
  \ee ^{-\sqrt{-z}\, \rho \cosh (u)-\ii \, \nu \, u}
  \left(\int _{-\infty }^{\infty }\ee ^{-\sqrt{-z}\, r\cosh (s)}
    \frac{\ee ^{-\alpha (u-s-\ii \theta )}}{1+\ee^{-u+s+\ii\theta}}\dd s
  \right)\dd u\, .
  \]
  Now it suffices to apply (\ref{eq:AsymptExpChFracExp}) to the inner
  bracket and then to use the integral form (\ref{eq:KimIntegralForm})
  in the reversed sense.
\end{proof}

First let us apply (\ref{eq:AsymptIntKK}) to the case of one vortex.
In fact, the following observation about the asymptotic expansion of
the Green function (\ref{eq:FinalGreenFceSepar}) near the vortex will
be crucial for the subsequent analysis. We get either the asymptotic
expansion of $\GG _{z}(r,\theta ;r_{0},\theta _{0})$ for
$r\downarrow0$, or, since in general it holds true that
\begin{equation}
  \overline{\GG _{\overline{z}}(r,\theta ;r_{0},\theta _{0})}
  =\GG _{z}(r_{0},\theta _{0};r,\theta ),
  \label{eq:GreenFceAdjoint}
\end{equation}
the expansion for $r_{0}\downarrow 0$ as well, namely
\begin{eqnarray}
  \GG _{z}(r,\theta ;r_{0},\theta _{0})
  & = & \frac{\sin (\pi \, \alpha )}{2\pi ^{2}}
  \frac{\Gamma (\alpha )}{1-\alpha }\left(
    \frac{\sqrt{-z}\, r_{0}}{2}\right)^{\! 1-\alpha }
  K_{-1+\alpha }\! \left(\sqrt{-z}r\right)\,
  \ee ^{\ii (\alpha -1)(\theta -\theta _{0})}
  \label{eq:AsymptGreenFceOneVort}\\
  &  & +\, \frac{\sin (\pi \, \alpha )}{2\pi ^{2}}
  \frac{\Gamma (1-\alpha )}{\alpha }\left(\frac{\sqrt{-z}\,
      r_{0}}{2}\right)^{\! \alpha }K_{\alpha }\!
  \left(\sqrt{-z}r\right)\, \ee ^{\ii \, \alpha (\theta -\theta _{0})}
  +O(r_{0})\,.\nonumber
\end{eqnarray}
One observes that the coefficients standing at
$r_{0}^{\,1-\alpha}\ee^{-\ii (\alpha -1)\theta _{0}}$ and
$r_{0}^{\,\alpha }\ee ^{-\ii \, \alpha \, \theta _{0}}$ are
respectively proportional to
\[
K_{-1+\alpha }\! \left(\sqrt{-z}\,
  r\right)\, \ee ^{\ii (\alpha -1)\theta }\textrm{ and }K_{\alpha }\!
\left(\sqrt{-z}\, r\right)\, \ee ^{\ii \, \alpha \, \theta }.
\]
But these functions are nothing but the basis functions in the
corresponding deficiency subspace, see
(\ref{eq:DeficiencyBasisOneVort}).

Next we shall consider the case of two vortices. Applying
(\ref{eq:AsymptIntKK}) to (\ref{eq:GreenFceTwoVort}) we
get
\begin{subequations}\label{eq:AsymptGreenFceTwoVort}
  \begin{eqnarray}
    \GG _{z}(x,x_{0}) & = & \frac{\sin (\pi \alpha )}
    {2\pi ^{2}}\frac{\Gamma (\alpha )}{1-\alpha }
    \bigg (\frac{\sqrt{-z}\, r_{a}}{2}\bigg )^{1-\alpha }
    \ee ^{\ii (\alpha -1)\theta _{a}}\LL _{\alpha -1}(x_{0})\nonumber \\
    &  & +\, \frac{\sin (\pi \, \alpha )}{2\pi ^{2}}
    \frac{\Gamma (1-\alpha )}{\alpha }\bigg (\frac{\sqrt{-z}\,
      r_{a}}{2}\bigg )^{\alpha }\ee ^{\ii \, \alpha \,
      \theta _{a}}\, \LL _{\alpha }(x_{0})+O(r_{a})
    \label{eq:AsymptGreenFceTwoVorta}
  \end{eqnarray}
  for $r_{a}\downarrow 0$ where
  \begin{eqnarray}
    &  & \hspace {-2em}\LL _{\nu }(x_{0})=K_{\nu }\!
    \left(\sqrt{-z}\, r_{0a}\right)\, \ee ^{-\ii \, \nu \,
      \theta _{0a}}+\sum _{\gamma ,\, n\geq 2,\, c_{n}=a}(-1)^{n-1}
    \int _{\mathbb{R}^{n-1}}K_{\ii \tau _{n-1}+\nu }\!
    \left(\sqrt{-z}\, \rho \right)\nonumber \\
    &  & \times \, K_{\ii (\tau _{n-2}-\tau _{n-1})}\!
    \left(\sqrt{-z}\, \rho \right)\times \ldots
    \times K_{\ii (\tau _{1}-\tau _{2})}\!
    \left(\sqrt{-z}\, \rho \right)K_{-\ii \tau _{1}}\!
    \left(\sqrt{-z}\, r_{0}\right)
    \nonumber \\
    &  & \times \, \frac{\sin (\pi \, \sigma _{n-1})}
    {\pi \sin (\pi (\sigma _{n-1}+\ii \tau _{n-1}))}
    \times \ldots \times \frac{\sin (\pi \, \sigma _{2})}
    {\pi \sin (\pi (\sigma _{2}+\ii \tau _{2}))}\,
    \frac{\sin (\pi \, \sigma _{1})\, \exp (-\theta _{0}\tau _{1})}
    {\pi \sin (\pi (\sigma _{1}+\ii \tau _{1}))}\, \dd ^{n-1}\tau
    \nonumber \\
    &  & \label{eq:AsymptGreenFceTwoVortb}
  \end{eqnarray}
\end{subequations}
(and, again, $(r_{0},\theta _{0})$ are the polar coordinates of the
point $x_{0}$ with respect to the center $c_{1}$). The convergence of
the series in (\ref{eq:AsymptGreenFceTwoVortb}) will be discussed
later in Section~\ref{sec:Deficiency-subspaces}.

\section{Deficiency subspaces for the case of two vortices
\label{sec:Deficiency-subspaces}}

In this section we are going to construct a basis in the deficiency
subspaces in the two-vortex case. So $H$ designates the two-vortex
spinless AB Hamiltonian described by the boundary conditions
(\ref{eq:BoundaryCondsTwoVort}), $X$ is the symmetric operator
obtained by restricting the domain of $H$ as described in
Section~\ref{sec:Asymptotic-behaviour} and $\NN (z)=\Ker (X^{*}-z)$ is
a deficiency subspace.

Asymptotic expansion (\ref{eq:AsymptGreenFceOneVort}) for the one
vortex case suggests that also in the two vortex case one may extract
from the Green function a basis in the deficiency subspace. From
(\ref{eq:AsymptGreenFceTwoVort}) and (\ref{eq:GreenFceAdjoint}) on
derives immediately a candidate for such a basis. It is formed by the
functions
\begin{subequations}\label{eq:DeficiencyBasisTwoVort}
  \begin{equation}
    \psi_{u,\nu ,z}(x) = \sum _{n=0}^{\infty }S_{n}(u,\nu ,z;x)\,,
    \label{eq:DeficiencyBasisTwoVorta}
  \end{equation}
  where
  \begin{equation}
    S_{0}(u,\nu ,z;x)=K_{\nu }\! \left(\sqrt{-z}\, r_{u}\right)\,
    \ee ^{\ii \, \nu \, \theta _{u}},
    \label{eq:DeficiencyBasisTwoVortb}
  \end{equation}
  \begin{eqnarray}
    &  & \hspace {-2em}S_{n}(u,\nu ,z;x)
    =(-1)^{n}\int _{\mathbb{R}^{n}}K_{\ii \tau _{n}}\!
    \left(\sqrt{-z}\, r_{n}\right)\label{eq:DeficiencyBasisTwoVortc}\\
    &  & \times \, K_{\ii (\tau _{n-1}-\tau _{n})}\!
    \left(\sqrt{-z}\rho \right)\times \ldots \times
    K_{\ii (\tau _{1}-\tau _{2})}\! \left(\sqrt{-z}\rho \right)
    K_{-\ii \tau _{1}-\nu }\! \left(\sqrt{-z}\rho \right)\nonumber \\
    &  & \times \, \frac{\sin (\pi \, \sigma _{n})\,
      \exp (\theta _{n}\, \tau _{n})}{\pi \sin (\pi (
      \sigma _{n}+\ii \tau _{n}))}\frac{\sin (\pi \, \sigma _{n-1})}
    {\pi \sin (\pi (\sigma _{n-1}+\ii \tau _{n-1}))}\times \ldots
    \times \frac{\sin (\pi \, \sigma _{1})}{\pi
      \sin (\pi (\sigma _{1}+\ii \tau _{1}))}\, \dd ^{n}\tau
    \nonumber
  \end{eqnarray}
  for $n\geq 1$, the indices are restricted to the range
  \begin{equation}
    u\in \{a,b\},\, \nu \in \{\omega -1,\omega \}\textrm{ where }
    \omega =\alpha \textrm{ if }u=a\textrm{, and }
    \omega =\beta \textrm{ if }u=b,
    \label{eq:DeficiencyBasisTwoVortd}
  \end{equation}
\end{subequations}
and to each $n\in \mathbb{N}$ one relates the unique alternating
sequence $(c_{n},c_{n-1},\ldots ,c_{1})$, $c_{j}\in \{a,b\}$ and
$c_{j}\neq c_{j+1}$, such that $c_{1}\neq u$. Correspondingly,
$\sigma_{j}=\alpha $ if $c_{j}=a$ and $\sigma _{j}=\beta $ if
$c_{j}=b$. As usual, $(r_{n},\theta _{n})=(r_{c_{n}},\theta _{c_{n}})$
are the polar coordinates with respect to the center $c_{n}$,
$(r_{c},\theta_{c})$ are the polar coordinates centered at the point
$c$.

Let us show that the series (\ref{eq:DeficiencyBasisTwoVorta}) actually
converges. In the Hilbert space $L^{2}(\mathbb{R},\dd \tau )$ we
introduce the vectors
\[
\gf _{u,z}(x;\tau )=K_{\ii \tau }\! \left(\sqrt{-z}\, r_{u}\right)
\exp(\theta _{u}\tau )\, \frac{\sin (\pi \, \sigma )}{\pi \,
  \sin (\pi(\sigma +\ii \tau ))},\textrm{ }\gg _{\nu ,z}(\tau )
=K_{-\ii \tau -\nu }\! \left(\sqrt{-z}\, \rho \right),
\]
and the operators $\gK _{z}$ and $\gD _{u}$ with the generalized
kernels
\[
\gK _{z}(\mu ,\omega )=K_{\ii (\omega -\mu )}\! \left(\sqrt{-z}\rho
\right),\textrm{ }\gD _{u}(\mu ,\omega )=\frac{
  \sin (\pi \, \sigma)}{\pi \, \sin (\pi (\sigma +\ii \mu ))}\,
\delta (\mu -\omega ),
\]
where
\[
u\in \{a,b\},\textrm{ }\sigma =\alpha \textrm{ if }u=a,
\textrm{ and }\sigma =\beta \, \, \textrm{if }\, \, u=b.
\]
For $u\in \{a,b\}$ let $v$ be the complementary vortex, i.e.,
$\{u,v\}=\{a,b\}$. For the sake of brevity we shall use the
matrix-like notation in the following paragraph. Thus the
transposition will in fact indicate an integration , i.e.,
$\gf^{T}\gg=\int_{\mathbb{R}}\gf(\tau)\,\gg(\tau)\dd \tau $.

We can rewrite the summands in equation
(\ref{eq:DeficiencyBasisTwoVorta}) using this notation (here
$n\geq1$),
\begin{eqnarray*}
  S_{2n-1}(u,\nu,z;x) &=& -\gf_{v,z}(x)^{T}
  (\gK_{z}\,\gD_{u}\,\gK_{z}\gD_{v})^{n-1}\gg_{\nu ,z}\,,\\
  S_{2n}(u,\nu,z;x) &=& \gf_{u,z}(x)^{T}\gK _{z}\,\gD_{v}
  (\gK_{z}\,\gD_{u}\,\gK_{z}\gD_{v})^{n-1}\gg_{\nu ,z}\,.
\end{eqnarray*}
These formulae make it possible to estimate the summands. Note that
$\gK _{z}$ acts as a convolution operator and so it is diagonalized by
the Fourier transform. Since
\[
\int _{-\infty }^{\infty }\ee ^{\ii \,
  x\, \tau }K_{\ii \tau }(a)\dd \tau =\pi \, \ee ^{-a\cosh (x)}
\]
we get
\[
\Vert \gK _{z}\Vert =\pi \, \ee ^{-\Re (\sqrt{-z})\rho }.
\]
The operator $\gD _{u}$ is already diagonal. Therefore
\[
\Vert \gD_{u}\Vert =\sup _{\mu \in \mathbb{R}}\left|\frac{\sin
    (\pi\sigma )}{\pi \sin (\pi (\sigma +\ii \mu
    ))}\right|=\frac{1}{\pi }\, .
\]
Jointly this implies that
\begin{eqnarray*}
  |S_{2n-1}(u,\nu,z;x)| &\leq& \|\gf_{v,z}(x)\|\|\gg_{\nu ,z}\|\,
  \ee^{-(2n-2)\,\Re(\sqrt{-z})\,\rho} \,,\\
  |S_{2n}(u,\nu,z;x)| &\leq& \|\gf_{u,z}(x)\|\|\gg_{\nu ,z}\|\,
  \ee^{-(2n-1)\,\Re(\sqrt{-z})\,\rho} \,.
\end{eqnarray*}
The estimates show that the series (\ref{eq:DeficiencyBasisTwoVorta})
converges absolutely at least as fast as a geometric series. Even one
can rewrite the formula for $\psi_{u,\nu ,z}(x)$ in a compact form,
namely
\begin{eqnarray}
  \psi _{u,\nu ,z}(x) & = & K_{\nu }\! \left(\sqrt{-z}\, r_{u}\right)
  \ee ^{\ii \, \nu \, \theta _{u}}
  \label{eq:PsiTwoVortSummed}\\
  &  & +\left(\gf_{u,z}(x)^{T}\gK _{z}\, \gD_{v}
    -\gf_{v,z}(x)^{T}\right)(\mathbb{I}-\gK_{z}\, \gD_{u}\,
  \gK_{z}\gD_{v})^{-1}\gg_{\nu ,z}.\nonumber
\end{eqnarray}
Here the inverse operator
$(\mathbb{I}-\gK_{z}\,\gD_{u}\,\gK_{z}\gD_{v})^{-1}$ exists with the
norm estimated from above by $(1-\exp(-2\Re(\sqrt{-z})\,\rho))^{-1}$.

Altogether we get four functions: $\psi _{a,\alpha -1,z}$,
$\psi_{a,\alpha ,z}$, $\psi _{b,\beta -1,z}$ and $\psi _{b,\beta ,z}$.
Our goal is to show that they actually form a basis in the deficiency
subspace. Obviously
\[
(\Delta +z)\psi _{u,\nu ,z}=0
\]
since
\[
(\Delta +z)K_{\nu }\! \left(\sqrt{-z}\, r\right)\ee ^{\pm \ii \, \nu
  \, \theta }=0\quad \textrm{on }\mathbb{R}^{2}\setminus \{0\}
\]
for all $\nu \in \mathbb{C}$, $z\in\mathbb{C}\setminus\mathbb{R}_{+}$,
and therefore all the summands satisfy the equation
$(\Delta+z)S_{n}(u,\nu ,z)=0$ in the domain
$\mathbb{R}^{2}\setminus(L_{a}\cup L_{b})$.

Let us verify that $\psi _{u,\nu ,z}$ obeys the boundary conditions
(\ref{eq:BoundaryCondsTwoVort}). For the sake of definiteness we shall
consider the function $\psi _{a,\nu ,z}$, $\nu\in\{\alpha-1,\alpha\}$.
Firstly we shall show that
\begin{equation}
  \ee ^{-\ii \, \alpha \, \pi }\psi _{a,\nu ,z}\big |_{\theta _{a}
    =\pi }-\ee ^{\ii \, \alpha \, \pi }\psi _{a,\nu ,z}
  \big |_{\theta _{a}=-\pi }=0.
  \label{eq:PsiVerifiesBCA}
\end{equation}
If $n=2m-1$ is odd then $c_{n}=b$. Moreover, if $\theta _{a}=\pm \pi $
then $\theta _{b}=0$ and $r_{b}=r_{a}+\rho $. Hence
\begin{eqnarray*}
  &  & \hspace {-1em}\ee ^{-\ii \, \alpha \, \pi }S_{2m-1}(a,\nu ,z)
  \big |_{\theta _{a}=\pi }-\ee ^{\ii \, \alpha \, \pi }
  S_{2m-1}(a,\nu ,z)\big |_{\theta _{a}=-\pi }\\
  &  & \quad =\, \int _{\mathbb{R}^{2m-1}}K_{\ii \tau _{2m-1}}\!
  \left(\sqrt{-z}(r_{a}+\rho )\right)\times \ldots \times
  K_{\ii (\tau _{1}-\tau _{2})}\! \left(\sqrt{-z}\, \rho \right)
  K_{-\ii \tau _{1}-\nu }\! \left(\sqrt{-z}\, \rho \right)\\
  &  & \qquad \times \, \frac{2\, \ii \sin (\pi \, \alpha )
    \sin (\pi \, \sigma _{2m-1})}{\pi \sin (\pi (\sigma _{2m-1}
    +\ii \tau _{2m-1}))}\times \ldots \times \frac{\sin (\pi \,
    \sigma _{1})}{\pi \sin (\pi (\sigma _{1}+\ii \tau _{1}))}\,
  {\dd ^{2m-1}}\tau \, .
\end{eqnarray*}
If $n=2m$, $m\geq 1$, is even then $c_{n}=a$ and
\[
\ee ^{-\ii \, \alpha \, \pi }\exp (\pi \, \tau _{n})-\ee ^{\ii \,
  \alpha \, \pi }\exp (-\pi \, \tau _{n})=-2\, \ii \,
\sin (\pi(\sigma _{n}+\ii \tau _{n}))
\]
hence
\begin{eqnarray*}
  &  & \hspace {-1em}\ee ^{-\ii \, \alpha \, \pi }S_{2m}(a,\nu ,z)
  \big |_{\theta _{a}=\pi }-\ee ^{\ii \, \alpha \, \pi }
  S_{2m}(a,\nu ,z)\big |_{\theta _{a}=-\pi }\\
  &  & \quad =-\int _{\mathbb{R}^{2m}}K_{\ii \tau _{2m}}\!
  \left(\sqrt{-z}\, r_{a}\right)K_{\ii (\tau _{2m-1}-\tau _{2m})}\!
  \left(\sqrt{-z}\, \rho \right)\times \ldots \times
  K_{-\ii \tau _{1}-\nu }\! \left(\sqrt{-z}\, \rho \right)\\
 &  & \qquad \times \, \frac{2\ii \, \sin (\pi \, \alpha )}{\pi }
 \frac{\sin (\pi \, \sigma _{2m-1})}{\pi \sin (\pi (\sigma _{2m-1}
   +\ii \tau _{2m-1}))}\times \ldots \times
 \frac{\sin (\pi \, \sigma _{1})}{\pi \sin (\pi (\sigma _{1}
   +\ii \tau _{1}))}\, \dd ^{2m}\tau \, .
\end{eqnarray*}
The integration in $\tau _{2m}$ can be carried on with the aid of the
identity
\begin{equation}
  \int _{-\infty }^{\infty }K_{\ii \tau }(a)\,
  K_{\ii (\nu -\tau )}(b)\dd \tau =\pi \, K_{\ii \nu }(a+b)\quad
  \textrm{for }a>0,\, b>0.
  \label{eq:ConvolutionKK}
\end{equation}
This way we get the equality
\begin{eqnarray*}
  &  & \ee ^{-\ii \, \alpha \, \pi }S_{2m}(a,\nu ,z)
  \big |_{\theta _{a}=\pi }-\ee ^{\ii \, \alpha \, \pi }
  S_{2m}(a,\nu ,z)\big |_{\theta _{a}=-\pi }\\
  &  & =-\big (\ee ^{-\ii \, \alpha \, \pi }S_{2m-1}(a,\nu ,z)
  \big |_{\theta _{a}=\pi }-\ee ^{\ii \, \alpha \, \pi }
  S_{2m-1}(a,\nu ,z)\big |_{\theta _{a}=-\pi }\big )
\end{eqnarray*}
valid for all $m\geq 1$. Obviously,

\[
\ee ^{-\ii \, \alpha \, \pi }S_{0}(a,\nu ,z)\big |_{\theta _{a}
  =\pi}-\ee ^{\ii \, \alpha \, \pi }S_{0}(a,\nu ,z)\big |_{\theta _{a}
  =-\pi}=0.
\]
The last two equalities imply (\ref{eq:PsiVerifiesBCA}).

Similarly one can show that
\begin{equation}
  \ee ^{-\ii \, \beta \, \pi }\psi _{a,\nu ,z}
  \big |_{\theta _{b}=\pi }-\ee ^{\ii \, \beta \, \pi }
  \psi _{a,\nu ,z}\big |_{\theta _{b}=-\pi }=0.
  \label{eq:PsiVerifiesBCB}
\end{equation}
Equality (\ref{eq:ConvolutionKK}) again turns out to be useful but
this time when treating the odd summands. With its aid the dimension
of the integration domain is reduced by 1 and one obtains the equality
\begin{eqnarray*}
  &  & \ee ^{-\ii \, \beta \, \pi }S_{2m-1}(a,\nu ,z)
  \big |_{\theta _{b}=\pi }-\ee ^{\ii \, \beta \, \pi }
  S_{2m-1}(a,\nu ,z)\big |_{\theta _{b}=-\pi }\\
  &  & =-\big (\ee ^{-\ii \, \beta \, \pi }S_{2m-2}(a,\nu ,z)
  \big |_{\theta _{b}=\pi }-\ee ^{\ii \, \beta \, \pi }
  S_{2m-2}(a,\nu ,z)\big |_{\theta _{b}=-\pi }\big )
\end{eqnarray*}
valid for all $m\geq 1$. This shows (\ref{eq:PsiVerifiesBCB}).

Finally we note that the remaining two boundary conditions,
\begin{eqnarray*}
  \ee ^{-\ii \, \alpha \, \pi }\, \frac{\partial
    \psi _{a,\nu ,z}}{\partial r_{a}}\Big |_{\theta _{a}=\pi }
  -\ee ^{\ii \, \alpha \, \pi }\,
  \frac{\partial \psi _{a,\nu ,z}}{\partial r_{a}}
  \Big |_{\theta _{a}=-\pi } & = & 0,\textrm{ }\\
  \ee ^{-\ii \, \beta \, \pi }\, \frac{\partial \psi _{a,\nu ,z}}
  {\partial r_{b}}\Big |_{\theta _{b}=\pi }
  -\ee ^{\ii \, \beta \, \pi }\, \frac{\partial \psi _{a,\nu ,z}}
  {\partial r_{b}}\Big |_{\theta _{b}=-\pi } & = & 0,
\end{eqnarray*}
can be verified in exactly the same way.

Next we wish to examine the asymptotic behavior of the functions
$\psi_{u,\nu ,z}$ near the singular points $a$ and $b$. We shall again
focus on the functions $\psi _{a,\nu ,z}$, the functions
$\psi_{b,\nu,z}$ can be treated similarly. First notice that
\begin{equation}
  \psi _{a,\nu ,z}(b)=0.\label{eq:PsiaAtbEq0}
\end{equation}
Actually, for the even summands in (\ref{eq:DeficiencyBasisTwoVorta})
the limit $x\to b$ just means setting $r_{a}=\rho $. To treat the odd
summands one applies the limit procedure (\ref{eq:LimitKtoDelta}) for
$r_{b}\rightarrow 0$ and finds that
\[
S_{2m-1}(a,\nu,z;b)=-S_{2m-2}(a,\nu ,z;b)\quad
\textrm{for all }m\geq 1.
\]
This shows (\ref{eq:PsiaAtbEq0}).

Let us make this result more precise. The even summands in
(\ref{eq:DeficiencyBasisTwoVorta}) simply satisfy
\[
S_{2m}(a,\nu ,z;x)=S_{2m}(a,\nu ,z;b)+O(r_{b})\quad \textrm{for }
x\, \rightarrow b.
\]
Asymptotic behavior of the odd summands can be obtained with the aid
of relation (\ref{eq:AsymptIntKK}). We get (here
$S_{2m-1}(a,\nu,z;b)=-S_{2m-2}(a,\nu ,z;b)$)
\begin{eqnarray*}
  &  & \hspace {-2em}S_{2m-1}(a,\nu ,z;x)=S_{2m-1}(a,\nu ,z;b)
  +\frac{\sin (\pi \, \beta )}{\pi }\frac{\Gamma (\beta )}{1-\beta }
  \bigg (\frac{\sqrt{-z}\, r_{b}}{2}\bigg )^{1-\beta }
  \ee ^{\ii (\beta -1)\theta _{b}}\\
  &  & \textrm{ }\times \int _{\mathbb{R}^{2m-2}}
  K_{\ii \tau _{2m-2}-1+\beta }\! \left(\sqrt{-z}\rho \right)
  \times \ldots \times K_{\ii (\tau _{1}-\tau _{2})}\!
  \left(\sqrt{-z}\, \rho \right)K_{-\ii \tau _{1}-\nu }\!
  \left(\sqrt{-z}\rho \right)\\
  &  & \textrm{ }\times \, \frac{\sin (\pi \, \sigma _{2m-2})}
  {\pi \sin (\pi (\sigma _{2m-2}+\ii \tau _{2m-2}))}\times\ldots\times
  \frac{\sin (\pi \, \sigma _{1})}{\pi
    \sin (\pi (\sigma _{1}+\ii \tau _{1}))}\, \dd ^{2m-2}\tau \\
  &  & \\
  &  & +\, \frac{\sin (\pi \, \beta )}{\pi }
  \frac{\Gamma (1-\beta )}{\beta }\bigg (\frac{\sqrt{-z}\, r_{b}}{2}
  \bigg )^{\beta }\ee ^{\ii \, \beta \, \theta _{b}}\\
  &  & \textrm{ }\times \int _{\mathbb{R}^{2m-2}}
  K_{\ii \tau _{2m-2}+\beta }\! \left(\sqrt{-z}\, \rho \right)
  \times \ldots \times K_{\ii (\tau _{1}-\tau _{2})}\!
  \left(\sqrt{-z}\, \rho \right)K_{-\ii \tau _{1}-\nu }\!
  \left(\sqrt{-z}\rho \right)\\
  &  & \textrm{ }\times \, \frac{\sin (\pi \, \sigma _{2m-2})}
  {\pi \sin (\pi (\sigma _{2m-2}+\ii \tau _{2m-2}))}
  \times \ldots \times \frac{\sin (\pi \, \sigma _{1})}
  {\pi \sin (\pi (\sigma _{1}+\ii \tau _{1}))}\, \dd ^{2m-2}\tau +O(r_{b})
\end{eqnarray*}
for $x\, \rightarrow b$. Jointly this means that
\begin{equation}
  \psi _{a,\nu ,z}(x)=\sum_{\mu\,\in\,\{\beta-1,\,\beta\}}
  \frac{\sin (\pi \left|\mu \right|)}{\pi }
  \frac{\Gamma (1-\left|\mu \right|)}{\left|\mu \right|}
  \left(\frac{\sqrt{-z}\, r_{b}}{2}\right)^{\! \left|\mu \right|}
  \ee ^{\ii \, \mu \, \theta _{b\, }}
  \sS _{\mu ,\nu }(\alpha ,\beta ;z)+O(r_{b})
  \label{eq:AsymPsiS}
\end{equation}
for $x\, \rightarrow b$ where
\begin{eqnarray}
  &  & \hspace {-3em}\sS _{\omega ,\nu }(\alpha ,\beta ;z)
  =K_{\omega -\nu }\! \left(\sqrt{-z}\rho \right)
  +\sum _{m=1}^{\infty }\int _{\mathbb{R}^{2m}}
  K_{\ii \tau _{2m}+\omega }\! \left(\sqrt{-z}\, \rho \right)\nonumber \\
  &  & \times \, K_{\ii (\tau _{2m-1}-\tau _{2m})}\!
  \left(\sqrt{-z}\, \rho \right)\times \ldots \times
  K_{\ii (\tau _{1}-\tau _{2})}\! \left(\sqrt{-z}\, \rho \right)
  K_{-\ii \tau _{1}-\nu }\! \left(\sqrt{-z}\, \rho \right)\label{eq:S}\\
  &  & \times \, \frac{\sin (\pi \, \sigma _{2m})}
  {\pi \sin (\pi (\sigma _{2m}+\ii \tau _{2m}))}\times \ldots \times
  \frac{\sin (\pi \, \sigma _{1})}{\pi
    \sin (\pi (\sigma _{1}+\ii \tau _{1}))}\, \dd ^{2m}\tau \nonumber
\end{eqnarray}
with
$(\sigma_{2m},\ldots,\sigma_{2},\sigma_{1})=(\alpha,\ldots,\alpha,\beta)$.

The function $\psi _{a,\nu ,z}$ has a singularity at the point $a$.
Nevertheless it holds true that
\begin{equation}
  \sum _{n=1}^{\infty}S_{n}(a,\nu ,z;a)=0.
  \label{eq:SumSaAta}
\end{equation}
The verification is similar to that of equality (\ref{eq:PsiaAtbEq0}).
This time it holds true that
\[
S_{2m}(a,\nu ,z;a)=-S_{2m-1}(a,\nu ,z;a)\quad \textrm{for all }m\geq
1.
\]
This shows (\ref{eq:SumSaAta}). A more precise result can be derived
as follows. Note that
\[
S_{2m-1}(a,\nu ,z;x)=S_{2m-1}(a,\nu ,z;a)+O(r_{a})\quad \textrm{for
}x\, \rightarrow a.
\]
Asymptotic behavior of the even summands can be obtained with the
aid of relation (\ref{eq:AsymptIntKK}). We get
\begin{eqnarray*}
  &  & \hspace {-2em}S_{2m}(a,\nu ,z;x)=S_{2m}(a,\nu ,z;a)
  -\frac{\sin (\pi \, \alpha )}{\pi }\frac{\Gamma (\alpha )}{1-\alpha}
  \bigg (\frac{\sqrt{-z}\, r_{a}}{2}\bigg )^{1-\alpha }
  \ee ^{\ii (\alpha -1)\theta _{a}}\\
  &  & \textrm{ }\times \int _{\mathbb{R}^{2m-1}}
  K_{\ii \tau _{2m-1}-1+\alpha }\! \left(\sqrt{-z}\, \rho \right)
  \times \ldots \times K_{\ii (\tau _{1}-\tau _{2})}\!
  \left(\sqrt{-z}\, \rho \right)K_{-\ii \tau _{1}-\nu }\!
  \left(\sqrt{-z}\, \rho \right)\\
  &  & \textrm{ }\times \, \frac{\sin (\pi \, \sigma _{2m-1})}
  {\pi \sin (\pi (\sigma _{2m-1}+\ii \tau _{2m-1}))}
  \times \ldots \times \frac{\sin (\pi \, \sigma _{1})}
  {\pi \sin (\pi (\sigma _{1}+\ii \tau _{1}))}\, \dd ^{2m-1}\tau \\
  &  & -\, \frac{\sin (\pi \, \alpha )}{\pi }
  \frac{\Gamma (\alpha )}{1-\alpha }\bigg (
  \frac{\sqrt{-z}\, r_{a}}{2}\bigg )^{1-\alpha }
  \ee ^{\ii \, \alpha \, \theta _{a}}\\
  &  & \textrm{ }\times \int _{\mathbb{R}^{2m-1}}
  K_{\ii \tau _{2m-1}+\alpha }\! \left(\sqrt{-z}\,
    \rho \right)\times \ldots \times
  K_{\ii (\tau _{1}-\tau _{2})}\! \left(\sqrt{-z}\, \rho \right)
  K_{-\ii \tau _{1}-\nu }\! \left(\sqrt{-z}\, \rho \right)\\
  &  & \textrm{ }\times \, \frac{\sin (\pi \, \sigma _{2m-1})}
  {\pi \sin (\pi (\sigma _{2m-1}+\ii \tau _{2m-1}))}
  \times \ldots \times \frac{\sin (\pi \, \sigma _{1})}
  {\pi \sin (\pi (\sigma _{1}+\ii \tau _{1}))}\, \dd ^{2m-1}\tau +O(r_{a})
\end{eqnarray*}
for $x\, \rightarrow a$. The asymptotic behavior of the Macdonald
function is given by the formula \cite{SpecFce}
\begin{eqnarray*}
  K_{\nu }(x) & = & \frac{\pi }{2\sin (\nu \, \pi )}
  \Big (\frac{2^{\nu }}{\Gamma (1-\nu )}\, x^{-\nu }-\frac{2^{-\nu }}
  {\Gamma (1+\nu )}\, x^{\nu }\Big )+O\big (x^{-\nu +2}\big )\\
  & = & \frac{\Gamma (\nu )}{2}\Big (\frac{x}{2}\Big )^{-\nu }
  -\frac{\Gamma (1-\nu )}{2\nu }\Big (\frac{x}{2}\Big )^{\nu }
  +O\big (x^{-\nu +2}\big )\quad \textrm{for }0<\nu <1.
\end{eqnarray*}
Finally we arrive at the expansion
\begin{eqnarray}
  \psi _{a,\nu ,z}(x) & = & \frac{\Gamma (\left|\nu \right|)}{2}
  \left(\frac{\sqrt{-z}\, r_{a}}{2}\right)^{\! -\left|\nu \right|}
  \ee ^{\ii \, \nu \, \theta _{a}}\label{eq:AsymPsiT}\\
  &  & -\, \sum _{\mu\,\in\,\{\alpha-1,\,\alpha\}}
  \frac{\sin (\pi \left|\mu \right|)}{\pi }
  \frac{\Gamma (1-\left|\mu \right|)}{\left|\mu \right|}
  \left(\frac{\sqrt{-z}\, r_{a}}{2}\right)^{\! \left|\mu \right|}
  \ee ^{\ii \, \mu \, \theta _{a}}\, \TT _{\mu ,\nu }(\alpha,\beta;z)
  + O(r_{a})\nonumber
\end{eqnarray}
for $x\, \rightarrow a$ where
\begin{eqnarray}
  &  & \hspace {-3em}\TT_{\omega ,\nu }(\alpha ,\beta ;z)
  = \frac{\pi }{2\sin (\pi \alpha)}\,
  \delta _{\mu \nu }+\sum _{m=1}^{\infty }
  \int _{\mathbb{R}^{2m-1}}K_{\ii \tau _{2m-1}+\omega }\!
  \left(\sqrt{-z}\, \rho \right)\nonumber \\
  &  & \hspace {-0.5em}\times \,
  K_{\ii (\tau _{2m-2}-\tau _{2m-1})}\!
  \left(\sqrt{-z}\, \rho \right)\times \ldots \times
  K_{\ii (\tau _{1}-\tau _{2})}\! \left(\sqrt{-z}\, \rho \right)
  K_{-\ii \tau _{1}-\nu }\! \left(\sqrt{-z}\, \rho \right)
  \label{eq:T}\\
  &  & \hspace {-0.5em}\times \, \frac{\sin (\pi \, \sigma _{2m-1})}
  {\pi \sin (\pi (\sigma _{2m-1}+\ii \tau _{2m-1}))}\times\ldots\times
  \frac{\sin (\pi \, \sigma _{1})}{\pi
    \sin (\pi (\sigma _{1}+\ii \tau _{1}))}\, \dd ^{2m-1}\tau \nonumber
\end{eqnarray}
with
$(\sigma_{2m-1},\ldots,\sigma_{2},\sigma_{1})=(\beta,\ldots,\alpha,\beta)$.

\begin{rem*}

As a consequence one can show that
\begin{equation}
  \sum _{n=1}^{\infty }S_{n}(u,\nu ,z;x)
  =\sum _{n=1}^{\infty }S_{n}(u,-\nu ,z;x).
  \label{eq:SEqS}
\end{equation}
Actually, a short inspection of the above derivation shows that the
functions
\[
\widetilde{\psi }_{u,\nu ,z}(x)
=\sum _{n=0}^{\infty }S_{n}(u,-\nu,z;x)
\]
also satisfy the boundary conditions (\ref{eq:BoundaryCondsTwoVort})
and solve the equation $(\Delta +z)\widetilde{\psi }_{u,\nu ,z}=0$.
Therefore the function
\[
f(x)=\sum _{n=1}^{\infty }S_{n}(u,\nu ,z;x)
-\sum _{n=1}^{\infty}S_{n}(u,-\nu ,z;x)
\]
satisfies the boundary conditions (\ref{eq:BoundaryCondsTwoVort}) as
well and solves $(\Delta +z)f=0$. In addition, $f(a)=f(b)=0$.
Consequently, $f\in \Dom (H)$ and $(H-z)f=0$. Necessarily $f=0$.

\end{rem*}

\begin{lem}\label{thm:DimDeficiSpaceLeq4}

$\dim \, \NN (z)\leq 4$.

\end{lem}

\begin{proof}

  In virtue of Corollary~\ref{thm:PsiFromDeficiSpaceGenSing}, for any
  five-tuple of functions from $\NN (z)$ there exists a nontrivial
  linear combination of these functions vanishing both at $a$ and $b$.
  By Corollary~\ref{thm:PsiDomHVanish}, such a linear combination
  equals 0.

\end{proof}

\begin{prop}

  $\dim \NN(z) = 4$.

\end{prop}

\begin{proof}

Owing to Lemma~\ref{thm:DimDeficiSpaceLeq4} it suffices to show
that $\dim \NN (z)\geq 4$. But in relation (\ref{eq:DeficiencyBasisTwoVort})
we have constructed four functions $\psi _{a,\alpha -1,z}$, $\psi _{a,\alpha ,z}$,
$\psi _{b,\beta -1,z}$ and $\psi _{b,\beta ,z}$ from the deficiency
subspace $\NN (z)$. The asymptotic expansions (\ref{eq:AsymPsiS})
and (\ref{eq:AsymPsiT}) show that these functions are actually linearly
independent.\end{proof}

We conclude that the functions $\{\psi _{a,\alpha -1,z},\psi _{a,\alpha ,z},\psi _{b,\beta -1,z},\psi _{b,\beta ,z}\}$
form a basis in $\NN (z)$.

\begin{figure}[hp]
  {\centering \hspace*{-5mm}\includegraphics[scale=0.85]{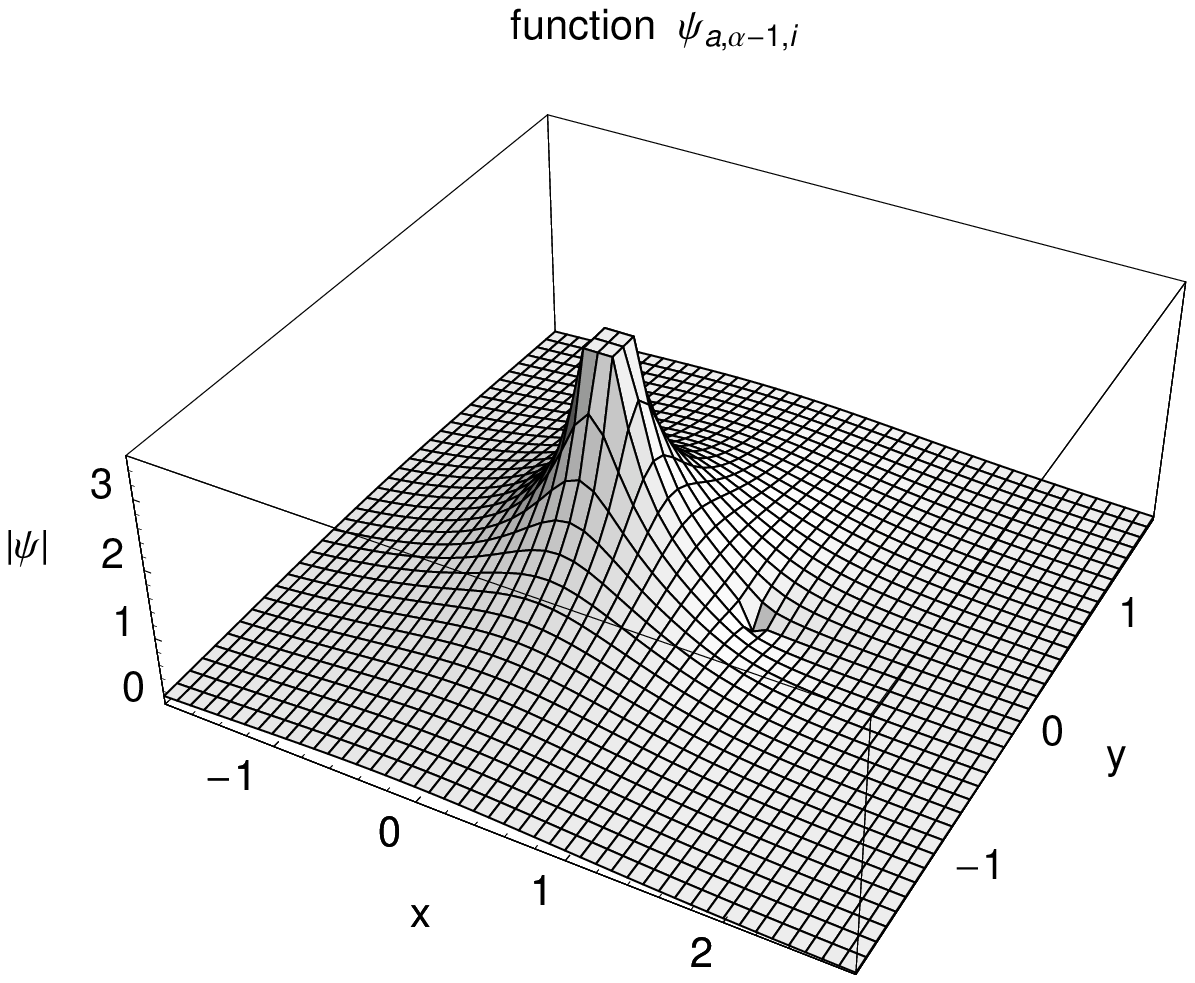}
    \par} \vskip 116pt
  \caption{Function $\psi_{a,\alpha-1,\mathrm{i}}$ from the deficiency
    subspace for the values of parameters $\alpha=1/3$, $\beta=2/3$,
    $\rho=1$.}
  \label{fig:psia}
\end{figure}

\begin{figure}[hp]
  {\centering \hspace*{-5mm}\includegraphics[scale=0.85]{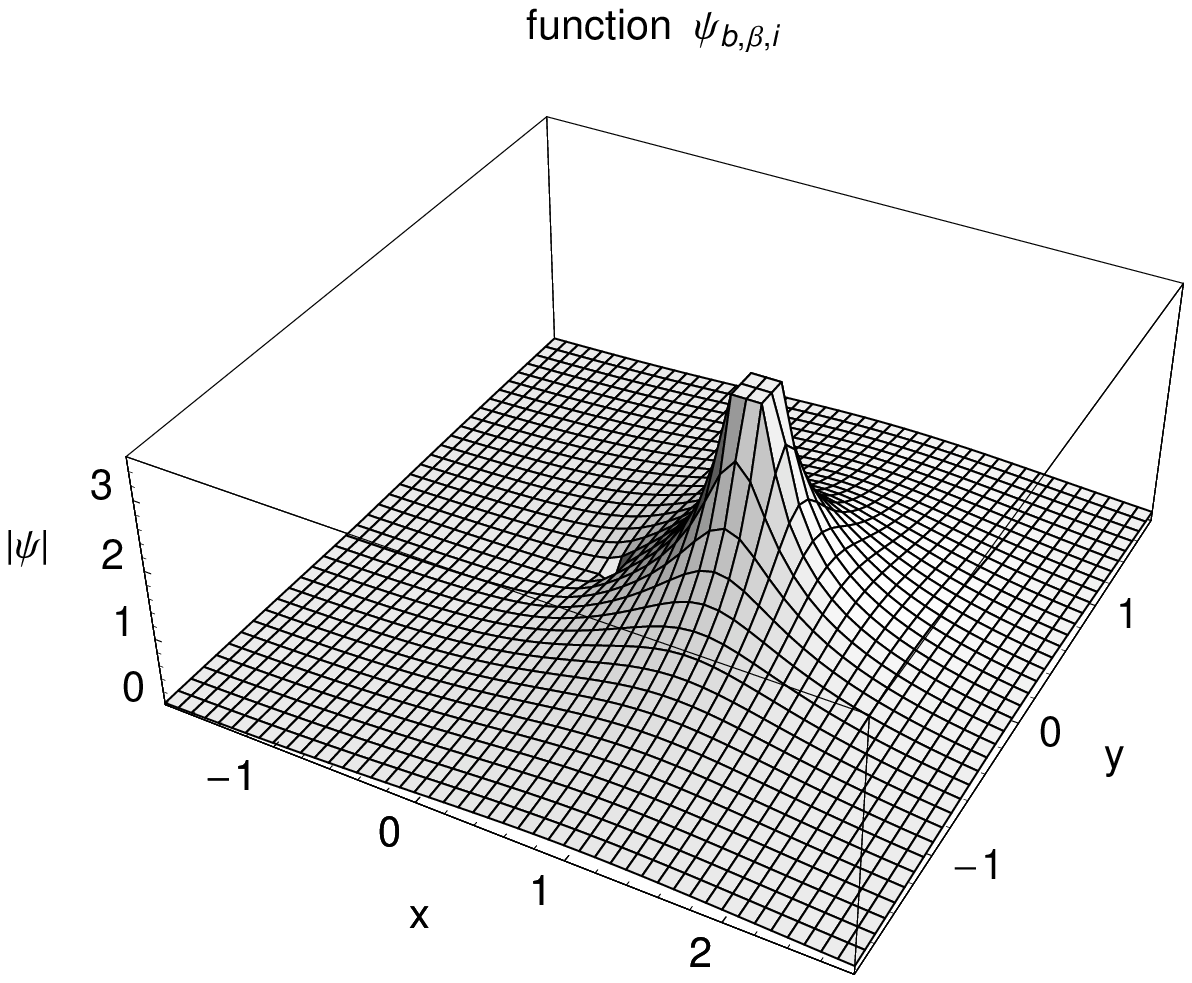}
    \par} \vskip 116pt
  \caption{Function $\psi_{b,\beta,\mathrm{i}}$ from the deficiency subspace
    for the values of parameters $\alpha=1/3$, $\beta=2/3$, $\rho=1$.}
  \label{fig:psib}
\end{figure}

\begin{rem*}
  
  Formula (\ref{eq:PsiTwoVortSummed}) is well suited for numerical
  computations. To give the reader an idea about the behavior of
  $\psi_{u,\nu ,z}$ we have plotted $|\psi _{a,\alpha -1,\ii }|$ in
  Fig.~\ref{fig:psia} and $|\psi _{b,\beta ,\ii }|$ in
  Fig.~\ref{fig:psib}, with $\alpha =1/3$, $\beta =2/3$ and $\rho =1$.
  Note that the former function vanishes in the vortex $b$ while the
  latter one vanishes in the vortex $a$.

\end{rem*}

\section{The Krein's formula\label{sec:The-Kreins-formula}}

We would like to emphasize once more that we are using two unitarily
equivalent formulations. The operators $H^{\pm }$, $H_{0}$ are
respectively associated to the quadratic forms (\ref{eq:QFormsQpm})
and (\ref{eq:QuadratFormTwoVort}).  Let $U$ be the unitary operator in
$L^{2}(\mathbb{R}^2,\dd^{2}x)$ acting as
$U\varphi=\exp(\ii\alpha\theta_{a}+\ii\beta\theta_{b})\varphi$.  The
Green function (\ref{eq:GreenFceTwoVort}) corresponds to the operator
$H=UH_{0}U^{-1}$ defined by the boundary conditions on the cut
(\ref{eq:BoundaryCondsTwoVort}).

Set\begin{equation}
f_{u,\nu ,z}=\Big (\sqrt{-z}\Big )^{\left|\nu \right|}\psi _{u,\nu ,z}.\label{eq:Deffunz}\end{equation}
Let us enumerate the basis $\{f_{a,\alpha -1,z},f_{a,\alpha ,z},f_{b,\beta -1,z},f_{b,\beta ,z}\}$
in $\NN (z)$ as $\{f_{z}^{1},f_{z}^{2},f_{z}^{3},f_{z}^{4}\}$ (in
this order). Set $\widetilde{f}_{z}^{j}=U^{-1}f_{z}^{j}$, $R_{z}=(H_{0}-z)^{-1}$,
$R_{z}^{\pm }=(H^{\pm }-z)^{-1}$. According to the Krein's formula\begin{equation}
R_{z}^{\pm }-R_{z}=\sum _{j,\, k}(M_{z}^{\pm })^{j,k}\, \widetilde{f}_{z}^{j}\langle \widetilde{f}_{\overline{z}}^{k},\cdot \rangle \label{eq:KreinsFormula}\end{equation}
or, in terms of Green functions,\begin{equation}
\GG _{z}^{\pm }(x,x_{0})=\GG _{z}(x,x_{0})+\sum _{j,\, k}(M_{z}^{\pm })^{j,k}\, f_{z}^{j}(x)\, \overline{f_{\overline{z}}^{k}(x_{0})}\, ,\label{eq:KreinsFormulaGreen}\end{equation}
where $M_{z}^{\pm }$ is a holomorphic matrix-valued function defined
on $\mathbb{C}\setminus \mathbb{R}$.

An operator-valued function $R_{z}^{\pm }$ constructed this way will
be the resolvent of a selfadjoint operator if and only if it satisfies
\cite[Chp. 5.2]{Weidmann}
\begin{equation}
  \forall z\in
  \mathbb{C}\setminus \mathbb{R},\textrm{ }(R_{z}^{\pm
  })^{*}=R_{\overline{z}}^{\pm }\label{eq:RpmAdjoint}
\end{equation}
and (the Hilbert identity)
\begin{equation}
  \forall z,w\in
  \mathbb{C}\setminus \mathbb{R},\textrm{ }R_{z}^{\pm }-R_{w}^{\pm
  }=(z-w)R_{z}^{\pm }R_{w}^{\pm }\,
  \label{eq:HilbertIdRpm}
\end{equation}
(it follows from (\ref{eq:KreinsFormula}) that $\Ker R_{z}^{\pm
}=\{0\}$ for all $z\in \mathbb{C}\setminus \mathbb{R}$). Let us
analyze conditions (\ref{eq:RpmAdjoint}) and (\ref{eq:HilbertIdRpm}).
It is straightforward to see that (\ref{eq:RpmAdjoint}) is satisfied
if and only if
\begin{equation}
  M_{z}^{*}=M_{\overline{z}}.
  \label{eq:MzAdjoint}
\end{equation}
In equality (\ref{eq:ResolventIdForfunz}) below we shall show that
\[
\forall z,w\in \mathbb{C}\setminus \mathbb{R},\, \forall j,\textrm{
}\widetilde{f}_{w}^{j}+(z-w)R_{z}\widetilde{f}_{w}^{j}=\widetilde{f}_{z}^{j}.
\]
With the aid of this identity it is just an easy computation to show
that (\ref{eq:HilbertIdRpm}) is equivalent to the
condition
\begin{equation}
  \forall z,w\in \mathbb{C}\setminus
  \mathbb{R},\textrm{ }M_{z}-M_{w}=(z-w)\, M_{z}P(\overline{z},w)\,
  M_{w}\,,
  \label{eq:IdentityForMandP}
\end{equation}
where $P(z,w)$ is the $4\times 4$ matrix of scalar products,
\[
P(z,w)^{j,k}=\langle f_{z}^{j},f_{w}^{k}\rangle .
\]
Equality (\ref{eq:IdentityForMandP}) was presented in
Ref.~\citen{AkhiezerGlazman} and was applied to problems similar to
ours for example in Refs.~\citen{DabrowskiGrosse} and
\citen{PSLDabrow}.

According to formula (\ref{eq:AsymptGreenFceTwoVort}) and definition
(\ref{eq:DeficiencyBasisTwoVort}) of $\psi _{u,\nu ,z}(x)$ we have
\begin{eqnarray}
  \GG _{z}(x,x_{0}) & = & \frac{\sin (\pi \, \alpha )}{2\pi ^{2}}
  \frac{\Gamma (\alpha )}{1-\alpha }\left(
    \frac{\sqrt{-z}\, r_{0a}}{2}\right)^{\! 1-\alpha }
  \ee ^{-\ii (\alpha -1)\theta _{0a}}\psi _{a,\alpha -1,z}(x)\nonumber \\
  &  & +\, \frac{\sin (\pi \, \alpha )}{2\pi ^{2}}
  \frac{\Gamma (1-\alpha )}{\alpha }\left(
    \frac{\sqrt{-z}\, r_{0a}}{2}\right)^{\! \alpha }
  \ee ^{-\ii \, \alpha \, \theta _{0a}}\, \psi _{a,\alpha ,z}(x)
  +O(r_{0a})
  \label{eq:AsymptGreenInPsi}
\end{eqnarray}
for $r_{0a}\downarrow 0$. Using this asymptotic behavior and the
Hilbert identity written in terms of Green functions,
\[
(z-w)\int_{\mathbb{R}^{2}}\GG _{z}(x,y)\, \GG _{w}(y,x_{0})
\dd^{2}y=\GG _{z}(x,x_{0})-\GG _{w}(x,x_{0})\, ,
\]
we obtain an equality valid for $u=a$, namely

\begin{equation}
  (z-w)\Big (\sqrt{-w}\Big )^{\left|\nu \right|}
  \int _{\mathbb{R}^{2}}\GG _{z}(x,y)\psi _{u,\nu ,w}(y)
  \dd ^{2}y=\Big (\sqrt{-z}\Big )^{\left|\nu \right|}
  \psi _{u,\nu ,z}(x)-\Big (\sqrt{-w}\Big )^{\left|\nu \right|}
  \psi _{u,\nu ,w}(x)\, .
  \label{eq:IntGreenPsi}
\end{equation}
This means that
\[
\psi_{u,\nu ,w}+(z-w)(H-z)^{-1}\psi_{u,\nu,w}
=\left(\frac{\sqrt{-z}}{\sqrt{-w}}
\right)^{\! \left|\nu\right|}\psi_{u,\nu ,z}
\]
for $\nu \in \{\alpha -1,\alpha \}$ and $u=a$. The same argument
naturally applies also to the vortex $u=b$. Using notation
(\ref{eq:Deffunz}) we find that
\begin{equation}
  f_{u,\nu,w}+(z-w)(H-z)^{-1}f_{u,\nu ,w}
  = f_{u,\nu,z}
  \label{eq:ResolventIdForfunz}
\end{equation}
holds true for all $w,z\in \mathbb{C}\setminus \mathbb{R}$.

We wish to compute the $4\times 4$ matrix of scalar products $P(z,w).$
Using (\ref{eq:GreenFceAdjoint}) and applying the asymptotic behavior
(\ref{eq:AsymptGreenInPsi}) once more, this time to equality
(\ref{eq:IntGreenPsi}), we find that the integral
\[
\int _{\mathbb{R}^{2}}\overline{\psi _{v,\mu ,z}(y)}\,
\psi _{u,\nu,w}(y)\dd ^{2}y
\]
equals the coefficient standing at
\[
\frac{\sin (\pi \left|\mu \right|)}{2\pi ^{2}}
\frac{\Gamma(1-\left|\mu \right|)}{\left|\mu \right|}
\Big (\frac{r_{v}}{2}\Big)^{\left|\mu \right|}
\ee ^{\ii \, \mu \, \theta _{v}}
\]
when taking the asymptotic expansion of the expression

\[
\frac{1}{\overline{z}-w}\left(
  \frac{1}{\big (\sqrt{-w}\big)^{\left|\nu \right|}}\,
  \psi _{u,\nu,\overline{z}}(x)
  -\frac{1}{\big (\sqrt{-\overline{z}}\big)^{\left|\nu \right|}}\,
  \psi _{u,\nu ,w}(x)\right)
\]
for $x\rightarrow v$, i.e., $r_{v}\downarrow 0$. In virtue of
(\ref{eq:AsymPsiT}) and (\ref{eq:AsymPsiS}) we get
\begin{eqnarray*}
  &  & \int _{\mathbb{R}^{2}}\overline{\psi _{a,\mu ,z}(y)}\,
  \psi _{a,\nu ,w}(y)\dd ^{2}y\\
  &  & \qquad \quad =-2\pi \, \frac{1}{\overline{z}-w}
  \left(\left(\frac{\sqrt{-\overline{z}}}
      {\sqrt{-w}}\right)^{\! \left|\nu \right|}
    \TT _{\mu ,\nu }(\alpha ,\beta ;\overline{z})
    -\left(\frac{\sqrt{-w}}{\sqrt{-\overline{z}}}\right)^{\!
      \left|\mu \right|}\TT _{\mu ,\nu }(\alpha ,\beta ;w)\right)
\end{eqnarray*}
and\begin{eqnarray*}
  &  & \int _{\mathbb{R}^{2}}\overline{\psi _{a,\mu ,z}(y)}\,
  \psi _{b,\nu ,w}(y)\dd ^{2}y\\
  &  & \qquad \quad =\, 2\pi \, \frac{1}{\overline{z}-w}
  \left(\left(\frac{\sqrt{-\overline{z}}}{\sqrt{-w}}\right)^{\!
      \left|\nu \right|}
    \sS _{\nu ,\mu }(\alpha ,\beta ;\overline{z})
    -\left(\frac{\sqrt{-w}}{\sqrt{-\overline{z}}}\right)^{\!
      \left|\mu \right|}\sS _{\nu ,\mu }(\alpha ,\beta ;w)\right).
\end{eqnarray*}
In particular,
\[
\int_{\mathbb{R}^{2}}\left|\psi _{a,\nu ,z}(x)\right|^{2}\, \dd^{2}x
= -\frac{2\pi}{\Im (z)}\, \Im \! \left(\left(\frac{\sqrt{-z}}
    {\sqrt{-\overline{z}}}\right)^{\!\left|\nu \right|}
  \TT _{\nu ,\nu }(\alpha ,\beta ;z)\right)\, .
\]
This means that, when passing to functions $\{f_{z}^{j}\}$ instead
of $\{\psi _{u,\nu ,z}\}$,
\begin{eqnarray}
  &  & \hspace {-4em}(\overline{z}-w)\, P(z,w)\nonumber \\
  &  & \textrm{ }\hspace {-2em}=-2\pi
  \left(\begin{matrix} \big (\sqrt{-\overline{z}}\big )^{2-2\alpha }
      \TT _{\alpha -1,\alpha -1}(\alpha ,\beta ;\overline{z}) &
      \sqrt{-\overline{z}}\,
      \TT _{\alpha ,\alpha -1}(\alpha ,\beta ;\overline{z})\\
      \sqrt{-\overline{z}}\,
      \TT _{\alpha -1,\alpha }(\alpha ,\beta ;\overline{z}) &
      \big (\sqrt{-\overline{z}}\big )^{2\alpha }
      \TT _{\alpha ,\alpha }(\alpha ,\beta ;\overline{z})\\
      -\big (\sqrt{-\overline{z}}\big )^{2-\alpha -\beta }
      \sS _{\alpha -1,\beta -1}(\beta ,\alpha ;\overline{z}) &
      -\big (\sqrt{-\overline{z}}\big )^{1+\alpha -\beta }
      \sS _{\alpha ,\beta -1}(\beta ,\alpha ;\overline{z})\\
      -\big (\sqrt{-\overline{z}}\big )^{1-\alpha +\beta }
      \sS _{\alpha -1,\beta }(\beta ,\alpha ;\overline{z}) &
      -\big (\sqrt{-\overline{z}}\big )^{\alpha +\beta }
      \sS _{\alpha ,\beta }(\beta ,\alpha ;\overline{z})
    \end{matrix}\right.\nonumber \\
  &  & \label{eq:Pzw}\\
  &  & \quad \left.\begin{matrix}
      -\big (\sqrt{-\overline{z}}\big )^{2-\alpha -\beta }
      \sS _{\beta -1,\alpha -1}(\alpha ,\beta ;\overline{z}) &
      -\big (\sqrt{-\overline{z}}\big )^{1-\alpha +\beta }
      \sS _{\beta ,\alpha -1}(\alpha ,\beta ;\overline{z})\\
      -\big (\sqrt{-\overline{z}}\big )^{1+\alpha -\beta }
      \sS _{\beta -1,\alpha }(\alpha ,\beta ;\overline{z}) &
      -\big (\sqrt{-\overline{z}}\big )^{\alpha +\beta }
      \sS _{\beta ,\alpha }(\alpha ,\beta ;\overline{z})\\
      \big (\sqrt{-\overline{z}}\big )^{2-2\beta }
      \TT _{\beta -1,\beta -1}(\beta ,\alpha ;\overline{z}) &
      \sqrt{-\overline{z}}\,
      \TT _{\beta ,\beta -1}(\beta ,\alpha ;\overline{z})\\
      \sqrt{-\overline{z}}\,
      \TT _{\beta -1,\beta }(\beta ,\alpha ;\overline{z}) &
      \big (\sqrt{-\overline{z}}\big )^{2\beta }
      \TT _{\beta ,\beta }(\beta ,\alpha ;\overline{z})
    \end{matrix}\right)\nonumber \\
  &  & \nonumber \\
  &  & -\, (\overline{z}\leftrightarrow w)\, .\nonumber
\end{eqnarray}

The Green functions $\GG _{z}^{\pm }(x,x_{0})$ should satisfy the
corresponding boundary conditions in each variable $x$, $x_{0}$.  Let
us first consider the case of $H^{+}$. Recall that the boundary
conditions which determine the domain of $H^{+}$ are
$\Phi_{2}^{-1}=\Phi _{1}^{0}=0$ (see (\ref{eq:BoundCondHplus})). Let
us check the asymptotic behavior of $\GG _{z}^{\pm }(x,x_{0})$ for
$x_{0}\rightarrow a$. Asymptotic behavior of $\GG _{z}(x,x_{0})$ is
given in (\ref{eq:AsymptGreenInPsi}) and asymptotic behavior of
$f_{z}^{j}(x_{0})$ follows from (\ref{eq:AsymPsiT}) and
(\ref{eq:AsymPsiS}) jointly with definition (\ref{eq:Deffunz}).  The
condition $\Phi _{1}^{0}=0$ means that the coefficient standing at
$(r_{0a}/2)^{-\alpha }\exp (-\ii \alpha \theta _{0a})$ vanishes.  This
term occurs only in the asymptotic expansion of $f_{z}^{2}(x_{0})$ and
so
\[
\sum _{j}(M_{z}^{+})^{j,2}\, f_{z}^{j}(x)=0\, .
\]
The set of functions $\{f_{z}^{j}\}$ is linearly independent and thus
we get a condition on the matrix $M_{z}^{+}$: $(M_{z}^{+})^{j,2}=0$
for all $j$. Considering the limit $x_{0}\rightarrow b$ one similarly
derives the condition $(M_{z}^{+})^{j,4}=0$. In view of
(\ref{eq:MzAdjoint}) one obtains more, namely
\begin{equation}
  (M_{z}^{+})^{j,k}=0\quad \textrm{whenever }j=2,4
  \textrm{ or }k=2,4.
  \label{eq:Mplus24Vanish}
\end{equation}
Let us denote by $M_{z}^{+,\textrm{red }}$ the reduced $2\times 2$
matrix obtained by omitting the vanishing rows and columns, i.e.,
\[
M_{z}^{+,\textrm{red }}=
\begin{pmatrix}
  (M_{z}^{+})^{1,1} & (M_{z}^{+})^{1,3}\\
  (M_{z}^{+})^{3,1} & (M_{z}^{+})^{3,3}
\end{pmatrix}.
\]

The condition $\Phi _{2}^{-1}=0$ for $x_{0}\rightarrow a$ means that
the coefficient standing at
$(r_{0a}/2)^{1-\alpha}\exp\!\big(-\ii(\alpha -1)\theta _{0a}\big )$
vanishes. Using (\ref{eq:Mplus24Vanish}) we get
\begin{eqnarray*}
  &  & \hspace {-3em}\frac{\sin (\pi \, \alpha )}{2\pi ^{2}}
  \frac{\Gamma (\alpha )}{1-\alpha }f_{z}^{1}(x)+\sum _{j}
  f_{z}^{j}(x)\bigg (-(M_{z}^{+})^{j,1}\,
  \frac{\sin (\pi \, \alpha )}{\pi }\frac{\Gamma (\alpha )}{1-\alpha }\\
  &  & \textrm{ }\times \, \big (\sqrt{-z}\, \big )^{2(1-\alpha )}\,
  \TT _{\alpha -1,\alpha -1}(\alpha ,\beta ;z)\\
  & & +\, (M_{z}^{+})^{j,3}\, \frac{\sin (\pi \, \alpha )}
  {\pi}\frac{\Gamma (\alpha )}{1-\alpha }
  \big (\sqrt{-z}\, \big)^{2-\alpha -\beta }\,
  \sS _{\alpha -1,\beta -1}(\beta ,\alpha;z)\bigg )=0.
\end{eqnarray*}
This is equivalent to the couple of equations
\begin{eqnarray*}
  \frac{1}{2\pi }-(M_{z}^{+})^{1,1}
  \big (\sqrt{-z}\, \big )^{2(1-\alpha )}\,
  \TT _{\alpha -1,\alpha -1}(\alpha ,\beta ;z)\hspace {12em} &  & \\
  +\, (M_{z}^{+})^{1,3}\big (\sqrt{-z}\, \big )^{2-\alpha -\beta }\,
  \sS _{\alpha -1,\beta -1}(\beta ,\alpha ;z) & = & 0,\\
  -\, (M_{z}^{+})^{3,1}\big (\sqrt{-z}\, \big )^{2(1-\alpha )}\,
  \TT _{\alpha -1,\alpha -1}(\alpha ,\beta ;z)\hspace {12em} &  & \\
  +\, (M_{z}^{+})^{3,3}\big (\sqrt{-z}\, \big )^{2-\alpha -\beta }\,
  \sS _{\alpha -1,\beta -1}(\beta ,\alpha ;z) & = & 0.
\end{eqnarray*}
Analogously, another two equations are obtained when considering the
limit $x_{0}\rightarrow b$, namely
\begin{eqnarray*}
  \frac{1}{2\pi }-(M_{z}^{+})^{3,3}\big (\sqrt{-z}\,\big)^{2(1-\beta)}\,
  \TT _{\beta -1,\beta -1}(\beta ,\alpha ;z)\hspace {12em} &  & \\
  +\, (M_{z}^{+})^{3,1}\big (\sqrt{-z}\, \big )^{2-\alpha -\beta }\,
  \sS _{\beta -1,\alpha -1}(\alpha ,\beta ;z) & = & 0,\\
  -\, (M_{z}^{+})^{1,3}\big (\sqrt{-z}\, \big )^{2(1-\beta )}\,
  \TT _{\beta -1,\beta -1}(\beta ,\alpha ;z)\hspace {12em} &  & \\
  +\, (M_{z}^{+})^{1,1}\big (\sqrt{-z}\, \big )^{2-\alpha -\beta }\,
  \sS _{\beta -1,\alpha -1}(\alpha ,\beta ;z) & = & 0.
\end{eqnarray*}
The four equations can be jointly rewritten in the matrix form,
\begin{eqnarray}
  &  & \hspace {-2em}M_{z}^{+,\textrm{red }}\label{eq:Mplusred}\\
  &  & \hspace {-0.8em}=\frac{1}{2\pi }
  \begin{pmatrix}
    \big (\sqrt{-z}\big )^{2-2\alpha }
    \TT _{\alpha -1,\alpha -1}(\alpha ,\beta ;z) &
    -\big (\sqrt{-z}\big )^{2-\alpha -\beta }
    \sS _{\beta -1,\alpha -1}(\alpha ,\beta ;z)\\
    -\big (\sqrt{-z}\big )^{2-\alpha -\beta }
    \sS _{\alpha -1,\beta -1}(\beta ,\alpha ;z) &
    \big (\sqrt{-z}\big )^{2-2\beta }
    \TT _{\beta -1,\beta -1}(\beta ,\alpha ;z)
  \end{pmatrix}^{-1}.\nonumber
\end{eqnarray}

It is straightforward to verify that the derived matrix $M_{z}^{+}$
actually obeys conditions (\ref{eq:MzAdjoint}) and (\ref{eq:IdentityForMandP}).
The former one follows from the equalities
\[
\overline{\TT _{\mu ,\nu }(\alpha ,\beta ;z)}
=\TT _{\mu ,\nu }(\alpha,\beta ;\overline{z}),\textrm{ }
\overline{\sS _{\mu ,\nu }(\alpha,\beta ;z)}
=\sS _{\mu ,\nu }(\alpha ,\beta ;\overline{z}),
\]
and
\[
\TT _{\mu ,\nu }(\alpha ,\beta ;z)
=\TT _{\nu ,\mu }(\alpha ,\beta;z),\textrm{ }
\sS _{\mu ,\nu }(\alpha ,\beta ;z)=\sS _{\nu ,\mu}(\beta ,\alpha ;z).
\]
The latter one follows from the form of $P(z,w)$ given in (\ref{eq:Pzw}).
In fact, (\ref{eq:Pzw}) and (\ref{eq:Mplusred}) jointly imply
\[
(\overline{z}-w)\, P(z,w)^{\textrm{red}}
= \big(M_{w}^{+,\textrm{red}}\big )^{-1}
-\big (M_{\overline{z}}^{+,\textrm{red}}\big)^{-1}.
\]

The other component of the Pauli operator, $H^{-}$, can be treated
similarly. The boundary conditions read $\Phi_{1}^{-1}=\Phi_{2}^{0}=0$
(see (\ref{eq:BoundCondHminus})). The condition $\Phi _{1}^{-1}=0$ for
$x_{0}\rightarrow a$ means that the coefficient standing at
$(r_{0a}/2)^{-1+\alpha }\exp \! \big (-\ii(\alpha-1)\theta_{0a}\big)$
vanishes. Hence
\[
\sum _{j}(M_{z}^{-})^{j,1}\, f_{z}^{j}(x)=0\, ,
\]
or equivalently, $(M_{z}^{-})^{j,1}=0$. Similarly for
$x_{0}\rightarrow 0$ we derive that $(M_{z}^{-})^{j,3}=0$, hence
\begin{equation}
  (M_{z}^{-})^{j,k}=0\quad \textrm{whenever }j=1,3\textrm{ or }k=1,3.
  \label{eq:Mminus13Vanish}
\end{equation}
Set
\[
M_{z}^{-,\textrm{red }}=
\begin{pmatrix}
  (M_{z}^{-})^{2,2} & (M_{z}^{-})^{2,4}\\
  (M_{z}^{-})^{4,2} & (M_{z}^{-})^{4,4}
\end{pmatrix}.
\]
The condition $\Phi _{2}^{0}=0$ for $x_{0}\rightarrow a$ means that
the coefficient standing at
$(r_{0a}/2)^{\alpha}\exp(-\ii\alpha\theta_{0a})$ vanishes. Using
(\ref{eq:Mminus13Vanish}) we get
\begin{eqnarray*}
  &  & \hspace {-5em}\frac{\sin (\pi \, \alpha )}{2\pi ^{2}}
  \frac{\Gamma (1-\alpha )}{\alpha }f_{z}^{2}(x)+\sum _{j}f_{z}^{j}(x)
  \bigg (-(M_{z}^{+})^{j,2}\, \frac{\sin (\pi \, \alpha )}{\pi }
  \frac{\Gamma (1-\alpha )}{\alpha }\\
  &  & \qquad \times \, \big (\sqrt{-z}\, \big )^{2\, \alpha }\,
  \TT _{\alpha ,\alpha }(\alpha ,\beta ;z)\\
  &  & \quad +\, (M_{z}^{+})^{j,4}\, \frac{\sin (\pi \,\alpha)}{\pi}
  \frac{\Gamma (1-\alpha )}{\alpha }
  \big (\sqrt{-z}\, \big )^{\alpha +\beta }\,
  \sS _{\alpha ,\beta }(\beta ,\alpha ;z)\bigg )=0.
\end{eqnarray*}
This is equivalent to the couple of equations
\begin{eqnarray*}
  \frac{1}{2\pi }-(M_{z}^{-})^{2,2}\big (\sqrt{-z}\,\big)^{2\alpha}\,
  \TT _{\alpha ,\alpha }(\alpha ,\beta ;z)+(M_{z}^{-})^{2,4}
  \big (\sqrt{-z}\, \big )^{\alpha +\beta }\,
  \sS _{\alpha ,\beta }(\beta ,\alpha ;z) & = & 0,\\
  -\, (M_{z}^{-})^{4,2}\big (\sqrt{-z}\, \big )^{2\alpha }\,
  \TT _{\alpha ,\alpha }(\alpha ,\beta ;z)+(M_{z}^{-})^{4,4}
  \big (\sqrt{-z}\, \big )^{\alpha +\beta }\,
  \sS _{\alpha ,\beta }(\beta ,\alpha ;z) & = & 0.
\end{eqnarray*}
For $x_{0}\rightarrow b$ one derives other two equations,
\begin{eqnarray*}
  \frac{1}{2\pi }-(M_{z}^{-})^{4,4}\big(\sqrt{-z}\, \big )^{2\beta}\,
  \TT _{\beta ,\beta }(\beta ,\alpha ;z)+(M_{z}^{-})^{4,2}
  \big (\sqrt{-z}\, \big )^{\alpha +\beta }\,
  \sS _{\beta ,\alpha }(\alpha ,\beta ;z) & = & 0,\\
  -(M_{z}^{-})^{2,4}\big (\sqrt{-z}\, \big )^{2\beta }\,
  \TT _{\beta ,\beta }(\beta ,\alpha ;z)+(M_{z}^{-})^{2,2}
  \big (\sqrt{-z}\, \big )^{\alpha +\beta }\,
  \sS _{\beta ,\alpha }(\alpha ,\beta ;z) & = & 0.
\end{eqnarray*}
Jointly the four equations mean that
\begin{equation}
  M_{z}^{-,\textrm{red }}=\frac{1}{2\pi }
  \begin{pmatrix}
    \big (\sqrt{-z}\big )^{2\alpha }
    \TT _{\alpha ,\alpha }(\alpha ,\beta ;z) &
    -\big (\sqrt{-z}\big )^{\alpha +\beta }
    \sS _{\beta ,\alpha }(\alpha ,\beta ;z)\\
    -\big (\sqrt{-z}\big )^{\alpha +\beta }
    \sS _{\alpha ,\beta }(\beta ,\alpha ;z) &
    \big (\sqrt{-z}\big )^{2\beta }
    \TT _{\beta ,\beta }(\beta ,\alpha ;z)
  \end{pmatrix}^{-1}.
  \label{eq:Mminusred}
\end{equation}

Let us note that the inverted matrices on the RHS of
(\ref{eq:Mplusred}) and (\ref{eq:Mminusred}) are actually well
defined. This is because the matrices in question depend on $z$
analytically in the domain $\mathbb{C}\setminus \mathbb{R}_{+}$ and
tend exponentially fast to invertible diagonal matrices for
$\Re\sqrt{-z}\rightarrow +\infty $ as one can easily deduce from the
discussion of the formula (\ref{eq:PsiTwoVortSummed}) related to the
convergence of the series (\ref{eq:DeficiencyBasisTwoVorta}) and from
the form of matrix entries (\ref{eq:S}) and (\ref{eq:T}).

\section{Concluding remarks}

Having a formula for the Green function $\GG _{z}^{\pm }(x,x_{0})$ it
would be, of course, desirable to use it for a more detailed analysis
of the Pauli operator, first of all for its spectral analysis. This
aim would assume, however, a more detailed analysis of the
functions $\sS _{\omega ,\nu }(\alpha ,\beta ;z)$ and
$\TT_{\omega,\nu}(\alpha ,\beta ;z)$. In particular, it would be
important to know what happens in the limit
$\Re\sqrt{-z}\rightarrow0$, i.e., when $z$ approaches
$\lambda\in\mathbb{R}_{+}$ from the upper or lower half-plane. Recall
that both $\sS _{\omega ,\nu }(\alpha ,\beta ;z)$ and
$\TT_{\omega,\nu}(\alpha ,\beta ;z)$ are expressed as infinite series
whose convergence is guaranteed for $\Re \sqrt{-z}>0$. Our first
attempts in this direction suggest that such an analysis might be
rather complex and should be considered as an independent problem in
its own right.

\bigskip

\renewcommand{\thesection}{\hskip -1em}
\section*{Acknowledgements}

\smallskip\noindent P.~\v{S}. wishes to acknowledge gratefully
a partial support from Grant No 201/01/0130 of the Grant Agency of
Czech Republic. V.~A.~G. is grateful to the Department of
Mathematics of Czech Technical University for the warm hospitality
and to INTAS (Grant No 00-257) and RFBR (Grant No
02-01-00804) for a financial support. The authors are
indebted to the referee for helpful comments.

\newpage
\begin{center}
\textbf{\Large Figure captions}
\end{center}
\vskip 12pt
\noindent
FIGURE 1. Geometrical arrangement. Choice of the cuts $L_a$, $L_b$ and
choice of the angle variables $\theta_a$, $\theta_b$.
\newline\vskip 24pt
\noindent
FIGURE 2. Function $\psi_{a,\alpha-1,\mathrm{i}}$ from the deficiency
subspace for the values of parameters $\alpha=1/3$, $\beta=2/3$,
$\rho=1$.
\newline\vskip 24pt
\noindent
FIGURE 3. Function $\psi_{b,\beta,\mathrm{i}}$ from the deficiency
subspace for the values of parameters $\alpha=1/3$, $\beta=2/3$,
$\rho=1$.
\newpage

\newpage

\newpage

\newpage

\begin{thebibliography}{10}
\bibitem{AB}Y. Aharonov, A. Bohm,
  "Significance of electromagnetic potentials in the quantum theory,"
  Phys. Rev. \textbf{115}, 485-491 (1959).
\bibitem{AharonovCasher}Y. Aharonov, A. Casher,
  "Ground state of spin 1/2 charged particle in a two dimensional
  magnetic field,"
  Phys. Rev. A \textbf{19}, 2461-2462 (1979).
\bibitem{PSLDabrow}L. Dabrowski, P. \v{S}\v{t}ov\'{\i}\v{c}ek,
  "Aharonov-Bohm effect with $\delta $-type interaction,"
  J.~Math. Phys. \textbf{39}, 47-62 (1998).
\bibitem{AdamiTeta}R. Adami, A. Teta,
  "On the Aharonov-Bohm Hamiltonian,"
  Lett. Math. Phys. \textbf{43}, 43-53 (1998).
\bibitem{PS:PLA}P. \v{S}\v{t}ov\'{\i}\v{c}ek,
  "The Green function for the two-solenoid Aharonov-Bohm effect,"
  Phys. Lett. A \textbf{142}, 5-10 (1989).
\bibitem{Ruijsenaars}S.~N.~M. Ruijsenaars,
  "The Aharonov-Bohm effect and scattering theory,"
  Ann. Phys. (Leipzig) \textbf{146}, 1-34 (1983).
\bibitem{Albeverioetal}S. Albeverio, F. Gesztesy, R. H\o{}egh-Krohn,
  H. Holden,
  "Point interactions in two dimensions. Basic properties,
  approximations and applications to solid state physics,"
  J. reine angew. Math. \textbf{380}, 87-107 (1987).
\bibitem{PS:withEV}P. Exner, P. Vyt\v{r}as,
  P. \v{S}\v{t}ov\'{\i}\v{c}ek,
  "Generalized boundary conditions for the Aharonov-Bohm effect
  combined with a homogeneous magnetic field,"
  J. Math. Phys. \textbf{43}, 2151-2168 (2002).
\bibitem{Ogurisu}O. Ogurisu,
  "Generalized boundary conditions of a spin-1/2 particle for
  the Aharonov-Bohm effect combined with a homogeneous magnetic field,"
  mp\_arc $02\textrm{-}495$.
\bibitem{Arai}A. Arai,
  "Properties of the Dirac--Weyl operator with a strongly singular
  gauge potential,"
  J. Math. Phys. \textbf{34}, 915-935 (1993).
\bibitem{GeylerGrishanov}V. A. Geyler, E. N. Grishanov,
  "Zero modes in a periodic system of Aharonov-Bohm solenoids,"
  JETP Letters \textbf{75}, 354-356 (2002).
\bibitem{SpecFce}A.~F. Nikiforov, V.~B. Uvarov,
  \textit{Special Functions of Mathematical Physics}
  (Birkh\"auser, Basel, 1988).
\bibitem{Weidmann}J. Weidmann,
  \emph{Linear Operators in Hilbert Spaces}
  (Springer-Verlag, New York, 1980).
\bibitem{AkhiezerGlazman}N.~I. Akhiezer, I.~M. Glazman,
  \textit{Theory of Linear Operators in Hilbert Spaces}
  (Pitman, London, 1981).
\bibitem{DabrowskiGrosse}L. Dabrowski, H. Grosse,
  "On nonlocal point interactions in one, two and three dimensions,"
  J. Math. Phys. \textbf{26}, 2777-2780 (1985).
\end{thebibliography}
\end{document}